\documentclass[twocolumn]{aastex631}

\usepackage{natbib}
\usepackage{enumerate}
\usepackage{multirow}

\newcommand{\orcid}[1]{\href{https://orcid.org/#1}{\includegraphics[width=10pt]{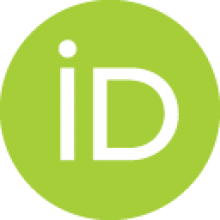}}}

\graphicspath{{./plot/}}

\begin{document}

\title{The Nature of Luminous Quasars with Very Large \ion{C}{4} Equivalent Widths}

\author[0000-0003-0964-7188]{Shuqi Fu}
\affiliation{CAS Key Laboratory for Research in Galaxies and Cosmology, Department of Astronomy, University of Science and Technology of China, Hefei 230026, China}
\affiliation{School of Astronomy and Space Sciences, University of Science and Technology of China, Hefei 230026, China}

\author[0000-0002-0167-2453]{W. N. Brandt}
\affiliation{Department of Astronomy and Astrophysics, 525 Davey Lab, The Pennsylvania State University, University Park, PA 16802, USA}
\affiliation{Institute for Gravitation and the Cosmos, The Pennsylvania State University, University Park, PA 16802, USA}
\affiliation{Department of Physics, 104 Davey Laboratory, The Pennsylvania State University, University Park, PA 16802, USA}

\author[0000-0002-4436-6923]{Fan Zou}
\affiliation{Department of Astronomy and Astrophysics, 525 Davey Lab, The Pennsylvania State University, University Park, PA 16802, USA}
\affiliation{Institute for Gravitation and the Cosmos, The Pennsylvania State University, University Park, PA 16802, USA}

\author[0000-0002-1615-179X]{Ari Laor}
\affiliation{Physics Department, Technion, Haifa 32000, Israel}

\author[0000-0002-7371-5416]{Gordon P. Garmire}
\affiliation{Huntingdon Institute for X-ray Astronomy, LLC, 10677 Franks Road, Huntingdon, PA 16652, USA}

\author[0000-0002-8577-2717]{Qingling Ni}
\affiliation{Institute for Astronomy, University of Edinburgh, Royal Observatory, Edinburgh EH9 3HJ, UK}

\author[0000-0001-8131-1801]{John D. Timlin III}
\affiliation{Department of Astronomy and Astrophysics, 525 Davey Lab, The Pennsylvania State University, University Park, PA 16802, USA}
\affiliation{Institute for Gravitation and the Cosmos, The Pennsylvania State University, University Park, PA 16802, USA}

\author[0000-0002-1935-8104]{Yongquan Xue}
\affiliation{CAS Key Laboratory for Research in Galaxies and Cosmology, Department of Astronomy, University of Science and Technology of China, Hefei 230026, China}
\affiliation{School of Astronomy and Space Sciences, University of Science and Technology of China, Hefei 230026, China}

\begin{abstract}

We report results for a complete sample of ten luminous radio-quiet quasars with large \ion{C}{4}
equivalent widths (EW $\geq 150$~\AA). For 8/10 we performed Chandra snapshot
observations. We find that, in addition to the enhanced \ion{C}{4} line EW, their \ion{He}{2} and 
\ion{Mg}{2} lines are enhanced, but the \ion{C}{3]} line is not. Their \mbox{X-ray} emission is substantially 
stronger than expected from their ultraviolet luminosity. Additionally, these large \ion{C}{4} EW quasars show 
small \ion{C}{4} blueshifts and possibly low Eddington ratios, suggesting they are ``extreme low 
Eigenvector 1 (EV1)" quasars.
The mean excess \ion{He}{2} EW is well-matched by Radiation Pressure Compression (RPC) 
photoionization models, with the harder $\alpha_{\mathrm{ox}}$ ionizing spectrum. However, these 
results do not reproduce well the enhancement pattern of the \ion{C}{4}, \ion{Mg}{2}, and \ion{C}{3]} 
EWs, or the observed high \ion{C}{4}/\ion{Mg}{2} ratio. RPC calculations indicate that the 
\ion{C}{4}/\ion{Mg}{2} line ratio is an effective metallicity indicator, and models with sub-Solar metallicity gas 
and a hard ionizing continuum reproduce well the enhancement pattern of all four ultraviolet lines. 
We find that the \ion{C}{4}/\ion{Mg}{2} line ratio in quasars is generally correlated with the excess \mbox{X-ray}
emission. Extremely high EV1 quasars are characterized by high metallicity and suppressed \mbox{X-ray}
emission. The underlying mechanism relating gas metallicity and \mbox{X-ray} emission is not clear, but 
may be related to radiation-pressure driven disk winds, which are enhanced at high metallicity, and
consequent mass loading reducing coronal \mbox{X-ray} emission.

\end{abstract}

\section{INTRODUCTION} \label{sec:intro}

Quasars are distinguishable from other astronomical sources by the typical presence of 
broad emission lines in their optical/ultraviolet (UV) spectra. These broad lines contain a 
wealth of information about the underlying physical properties of quasars. In particular, the 
\ion{C}{4} broad emission line in luminous quasars has been the target of extensive 
research. \ion{C}{4}$\ \lambda1549$ rest-frame equivalent width (\ion{C}{4} EW) and 
blueshift are known to correlate with the UV--\mbox{X-ray} power-law slope 
($\alpha_{\mathrm{ox}}$;\footnote{$\alpha_{\mathrm{ox}} = 0.3838 \times\mathrm{log}_{10}(f_{2\mathrm{keV}}/f_{2500})$, where $f_{2\mathrm{keV}}$ and $f_{2500}$ are the flux density at rest-frame 2 keV and 2500~\AA, respectively.} e.g.,\ \citealt{Green1998,Gibson2008,Richards2011,Timlin2020}) 
and to be linked to the Eddington ratio (i.e., the ratio between the bolometric luminosity and 
the Eddington luminosity; e.g.,\ \citealt{Baskin2004,Shen2014,Rivera2020}). Additionally, 
\mbox{X-ray} studies of Sloan Digital Sky Survey (SDSS; \citealt{York2000}) quasars have 
confirmed a positive correlation between \ion{C}{4} EW and $\Delta\alpha_{\mathrm{ox}}$,
\footnote{$\Delta\alpha_{\mathrm{ox}} = \alpha_{\mathrm{ox}} - \alpha_{\mathrm{ox}}(L_{2500})$, where $\alpha_{\mathrm{ox}}(L_{2500})$ is the expected $\alpha_{\mathrm{ox}}$ at a specified value of $L_{2500}$ from the relation in \citet{Just2007}.} 
a parameter quantifying the strength of the \mbox{X-ray} emission compared to that of a 
typical quasar matched in UV luminosity (\citealt{Gibson2008, Timlin2020}). Physically, this 
correlation is likely expected since there are more \mbox{\mbox{X-ray}/EUV} (extreme UV) 
ionizing photons to produce \ion{C}{4} ions in quasars with stronger \mbox{X-ray} emission. 
The quasars in these correlation studies have \ion{C}{4} EWs largely in the range 20--100~\AA\ 
due to the rarity of quasars outside this \ion{C}{4} EW range. Survey observations, 
however, have found a small population of luminous radio-quiet quasars that have very large \ion{C}{4} 
EWs (\ion{C}{4} EW $\gtrsim 150$~\AA) as well as radio-quiet quasars with extraordinarily weak lines 
\mbox{(\ion{C}{4} EW $\lesssim 15$~\AA)}. \mbox{X-ray} observations of weak-line quasars 
(WLQs) continue to provide insights into the nature of these objects; however, the 
\mbox{X-ray} and other properties of luminous quasars with very large \ion{C}{4} EWs have not been well studied. 

WLQs are a prime example of how these rare populations exhibit extraordinary 
\mbox{X-ray} properties. For example, nearly half of WLQs have been observed to be 
\mbox{X-ray} weak ($\Delta\alpha_{\mathrm{ox}} \leq -0.2$; e.g., \citealt{Luo2015,Ni2018,Ni2022}), many of which 
are weak by factors of $\approx$ 20--80 or more with respect to typical quasars (e.g., \citealt{Just2007}), and 
appear to be significantly \mbox{X-ray} absorbed. The other half have nominal-strength \mbox{X-ray} emission ($\Delta\alpha_{\mathrm{ox}} \approx 0$). WLQs appear to have large Eddington ratios as indicated by their steeper intrinsic X-ray power-law continua with respect to typical quasars \citep{Luo2015, Marlar2018}. 
The high incidence of \mbox{X-ray} weakness among WLQs is not observed in the general quasar population. 
As part of their work, \citet{Ni2018,Ni2022} generated a representative, unbiased sample of 32 WLQs (and 63 total 
WLQs) to investigate further the \mbox{$\Delta\alpha_{\mathrm{ox}}$ -- \ion{C}{4} EW} relation and found that the 
WLQs exhibit a strikingly large dispersion around the best-fit trend derived for typical quasars. \mbox{X-ray} 
observations of the rare population of quasars with very large \ion{C}{4} EWs might find similarly notable properties 
and furthermore help to constrain better the \mbox{$\Delta\alpha_{\mathrm{ox}}$ -- \ion{C}{4} EW} correlation.

To increase the dynamic range of \ion{C}{4} EW coverage and improve the significance of the 
\mbox{$\Delta\alpha_{\mathrm{ox}}$ -- \ion{C}{4} EW} correlation, \citet{Timlin2020} (hereafter T20) 
searched for serendipitous Chandra observations of quasars in the SDSS fourteenth data release 
(DR14Q, \citealt{Paris2018}). Quality cuts were imposed on the data (e.g., cuts in redshift and on sensitivity of 
the Chandra observations) to generate an unbiased sample of quasars with a high \mbox{X-ray} detection 
fraction. The final ``Sensitive'' sample contains 753 typical quasars with \mbox{X-ray} flux measurements, 
637 of which also have \ion{C}{4} EW measurements (hereafter the ``\ion{C}{4} subsample''). 
T20 found that the strength of the ionizing emission remains positively correlated with \ion{C}{4} EW even after 
accounting for the dependence of $\alpha_{\rm ox}$ on $L_{2500}$ (represented by $\Delta\alpha_{\mathrm{ox}}$). 
However, the T20 serendipitous sample included few quasars with very large \ion{C}{4} EWs ($\geq$~150~\AA), 
making it unclear if the correlation fitted to the typical quasars can be appropriately extrapolated to larger 
\ion{C}{4} EWs. 

Many of the quasars with very large \ion{C}{4} EWs ($\rm 150-185~\AA$) included in the T20 sample are much less 
luminous than the full sample and currently studied WLQs. Large \ion{C}{4} EWs are rarely found for luminous 
quasars due to the observed anti-correlation between quasar continuum luminosity and \ion{C}{4} EW (the Baldwin 
effect; \citealt{Baldwin1977}). Among the 637 quasars in the T20 sample, only two have large \ion{C}{4} EWs 
(\ion{C}{4} EW $\gtrsim 150$~\AA) and relatively large rest-frame 2500~\AA\ luminosity 
($L_{\rm 2500} \gtrsim 10^{30.5} \ \rm erg\ s^{-1}\ Hz^{-1}$). These two quasars differ in \mbox{X-ray} strength 
(measured by $\alpha_{\mathrm{ox}}$) by a factor of $\approx3.5$, making it difficult to determine if this outlier 
population has \mbox{X-ray} properties consistent with the general quasar population or if it has notable properties 
analogously to WLQs. We therefore proposed Chandra snapshot observations of eight quasars with large 
\ion{C}{4} EWs to understand better the relation between their \mbox{X-ray} continuum and UV emission-line 
strengths. 

Studies of such extreme objects can also help to clarify the dependence of quasar broad line emission upon factors including ionizing continuum strength and metallicity. 
For example, photoionization calculations have shown that while the strengths of two well-studied high-ionization emission lines, \ion{C}{4} and \ion{He}{2}$\ \lambda1640$, both depend upon the strength of the EUV ionizing radiation, they have different sensitivities to the metallicity of the quasar broad emission-line region gas (e.g., see Figure 5 of \citealt{Baskin2014}). 
Higher metallicity cools the gas, and weakens the \ion{C}{4} line. The \ion{He}{2} line, on the other hand, provides a relatively ``clean" measure of the number of ionizing photons, and is nearly independent of metallicity. 
The targeted quasars with large \ion{C}{4} EW should allow us to test this and related behavior over a wide range of parameters.

This paper is organized as follows. In Section 2 we describe the sample selection, our Chandra observations 
and data reduction, and the optical spectral-fitting methods. In Section 3 we show our main findings regarding the 
distinct optical and \mbox{X-ray} properties of our target quasars. Possible explanations of their observed properties 
are given in Section 4. Finally, Section 5 summarizes the results. Throughout this work, we adopt a flat 
\mbox{$\Lambda$-CDM} cosmology with $H_0=70\ \rm km\ s^{-1}\ Mpc^{-1}$, $\Omega_M=0.3$, and 
$\Omega_\Lambda=0.7$.

\section{SAMPLE SELECTION AND DATA} \label{sec:data}

\subsection{Sample selection} \label{subsec:sample}

\begin{figure*}[!t]
\gridline{\fig{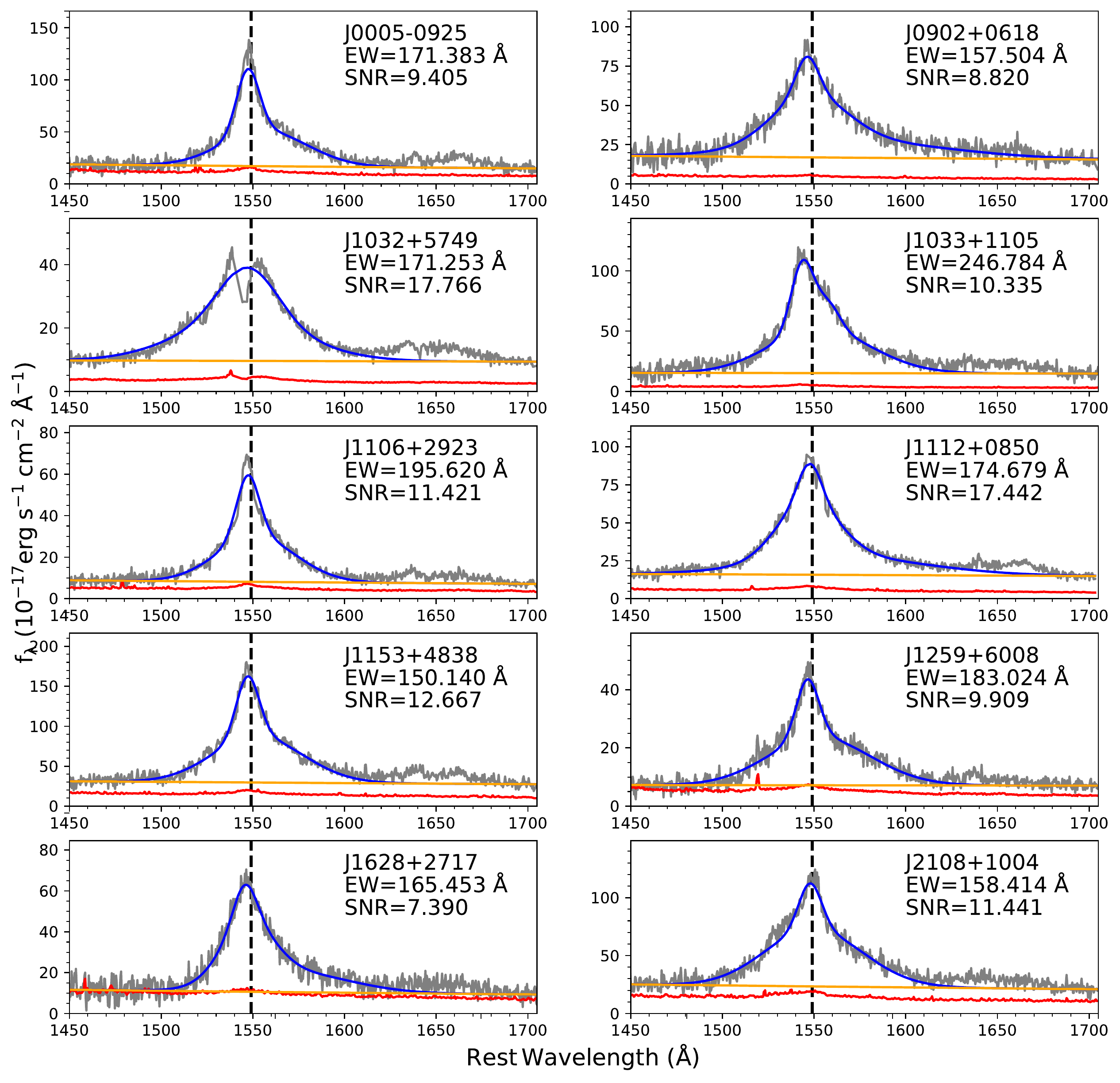}{0.5\textwidth}{(a)}
          \fig{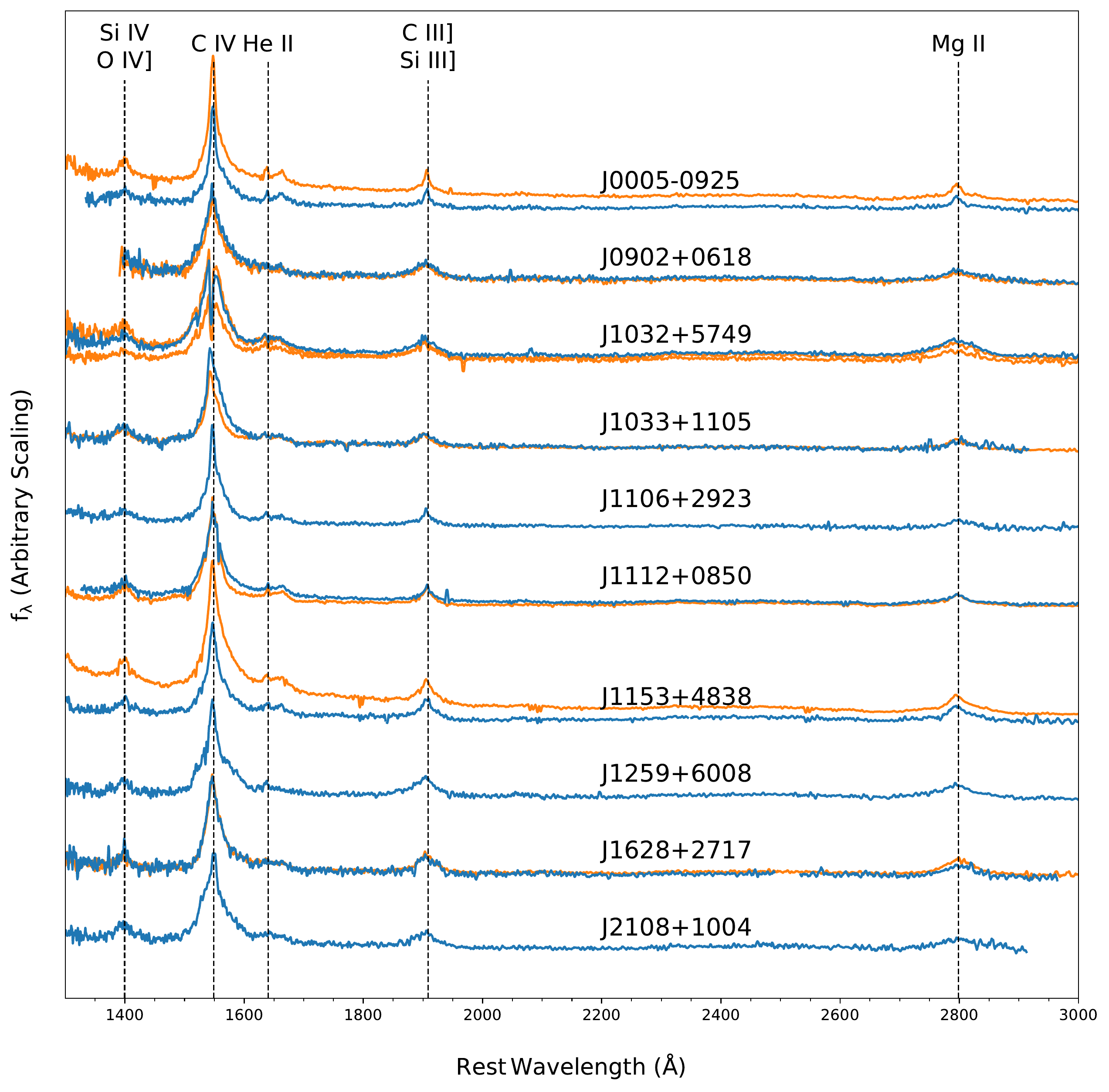}{0.5\textwidth}{(b)}}
\caption{(a) The \ion{C}{4} line region of the SDSS DR7 spectra of the eight target quasars and two large \ion{C}{4} EW quasars from T20 (SDSS \mbox{J103215.88+574926.4} and SDSS \mbox{J125929.13+600846.0}) included in our sample. The grey, red, blue, and orange lines are data, error, three broad Gaussian profiles, and local power-law continuum, respectively. The vertical dashed lines show the laboratory wavelengths of the \ion{C}{4} line ($\lambda \rm = 1549.06~\AA$). The rest-frame equivalent width ($\rm \AA$) of the \ion{C}{4} line measured from the Gaussian models and the median signal-to-noise ratio (SNR) per pixel for the \ion{C}{4} line are listed in the upper right. These \ion{C}{4} lines have high spectral quality ($\rm SNR>7$) and large rest-frame equivalent widths ($\rm EW>150\ \AA$).  
(b)  All the available SDSS spectra of the ten large \ion{C}{4} EW quasars, shown in the rest-frame. The visible UV emission lines are marked with dashed lines. The SDSS DR7 spectra are shown in blue, while other observations, if any, are shown in orange.
\label{fig1}}
\end{figure*}

We selected our targets from the \citet{Shen2011} quasar catalog, which contains 17 quasars that have \ion{C}{4} EW $\geq 150$~\AA. We also require our targets to be optically bright ($i$-band magnitude $< 20$) and not significantly red compared to typical quasars or WLQs ($\Delta(g-i) \leq 0.15$\footnote{Relative color, defined by subtracting the median colors of quasars at the redshift of each quasar from the measured colors of each quasar.}; see \citealt{Richards2003}), and to have no previous sensitive \mbox{X-ray} coverage. A total of eight quasars were selected with redshifts ranging from \mbox{1.7--2.2}.  We adopted the improved redshift measurements from \citet{Hewett_Wild2010}, based on cross-correlation (with a master quasar template) that includes \ion{Mg}{2}$\ \lambda2799$ or \ion{C}{3}$]\ \lambda1908$, as the systemic redshifts for the target quasars, which have systematic biases a factor of $\approx 20$ lower compared to the SDSS pipeline redshift values. We also inspected their SDSS spectra visually, and no broad absorption lines are present. Our targets are also more luminous (with a median $M_i=-26.4$; see Table~\ref{tab1}) than the quasars in the T20 sample that have comparable \ion{C}{4} EWs except for the two with similar UV luminosities mentioned in Section \ref{sec:intro}. These two, SDSS \mbox{J103215.88+574926.4} and SDSS \mbox{J125929.13+600846.0}, will be included in the following analyses due to their similarity to our new sources (see Section~\ref{sec:prop}). The \ion{C}{4} line regions of the SDSS DR7 spectra of the ten quasars are shown in Figure~\ref{fig1}a, and the whole spectra are shown in Figure~\ref{fig1}b. 

These ten quasars are observed to be radio quiet in the Very Large Array FIRST survey (\citealt{Becker1995}). We derived 2$\sigma$ upper limits on the 20~cm radio flux as $0.25 + 2\sigma_{\rm rms}$ mJy, where $\sigma_{\rm rms}$ is the RMS flux at the source position and 0.25~mJy is the CLEAN bias correction (\citealt{White1997}),  and we obtain flux densities at rest-frame \mbox{2500~\AA\ ($f_{2500}$)} by converting $M_i$($z = 2$) (\citealt{Richards2006}) to monochromatic luminosity at \mbox{2500~\AA\ ($L_{\rm 2500}$)} and then converting $L_{\rm 2500}$ to $f_{2500}$.\footnote{$L_{\rm 2500} = 4\pi D^2_L f_{2500}/(1 + z)$, where $D_L$ is the luminosity distance.} Then we calculate the upper limits on the radio-loudness parameter $R$,\footnote{The radio-loudness parameter is defined as the ratio between the 6 cm flux density and the 2500 $\rm \AA$ flux density: $R = f_{6cm}/f_{2500}$ (\citealt{Richards2011}).} and all of our targets have $R < 5$. Thus no additional strong \mbox{X-ray} emission is expected due to quasar jets or jet-linked enhanced coronae (e.g., \citealt{Zhu2020,Zhu2021}). 

The photometric properties used in the sample selection are listed in Table \ref{tab1}.

\subsection{Chandra observation and data reduction} \label{subsec:chandra}
We performed Chandra snapshot observations of the selected eight quasars with large \ion{C}{4} EWs. Chandra ACIS is the ideal instrument to carry out these observations due to its high sensitivity and low background, compared to \emph{XMM-Newton}, for snapshot point-source observations. The exposure times of our targets range from 2.8 to 6.0 ks. The observation details are summarized in Table \ref{tab2}.
    
We used standard CIAO tools to reduce our \mbox{X-ray} data, following the procedures described in section~3 of T20. After the ${\tt\string chandra\_repro}$ processing, we extracted a light curve, and then used the {\tt\string deflare} tool to remove any time intervals with background flares above the $3\sigma$ level in the light curve. Then we employed the {\tt\string fluximage} tool to generate images and exposure maps for both the soft (\mbox{0.5--2 keV}) and hard (\mbox{2--7 keV})  bands. After the images were created, {\tt\string wavdetect} was run on each image to search for sources. For each detected source, we compared the positions from {\tt\string wavdetect} with the SDSS coordinates, and if the coordinates of a detected source in either band matched within $0.5\arcsec$ of the SDSS position, we adopted this coordinate as the source position.  Around this position, we use {\tt\string srcflux} to create a circular region that encloses $90\%$  of the PSF at 1.0 keV, and then we extracted the raw counts within the circular region with the radius plus 5 additional pixels (e.g., \citealt{Gibson2008}). The background was extracted in an annulus with radius equal to the source radius plus 15 (50) pixels for the inner (outer) radius. For \mbox{J103312.84+110555.3} there is another faint source in the hard band in the background region, and we removed it manually. We visually inspected our source and background regions and verified that there are no other apparent extraneous sources in these regions. We calculated the $1\sigma$ errors of the net counts, based on the Poisson errors on the extracted source and background counts (\citealt{Gehrels1986}). The effective exposure time corrected for vignetting in both the source and background regions can be derived from the exposure maps. Finally, we used the {\tt\string specextract} tool to create the source and background spectra. 

As in T20, we computed the band ratios (i.e., the ratio of the hard-band to soft-band counts) with uncertainties using the Bayesian Estimation of Hardness Ratio (BEHR; \citealt{Park2006}) package. To compare with the T20 quasar sample, we first used the same method as theirs to estimate the effective power-law photon index ($\Gamma_{\rm eff}$), i.e., using the {\tt\string modelflux} tool, given the instrumental responses and a model of a power-law with Galactic absorption. We minimized the difference between the measured hard-band count rate and the prediction from the soft-band count rate and a series of $\Gamma$ using {\tt\string modelflux} to determine $\Gamma_{\rm eff}$. We then estimate the rest-frame flux density at 2~keV, $f_{\rm 2keV}$, using $\Gamma_{\rm eff}$ and the measured soft-band count rate. We deduced the upper and lower limits for $f_{\rm 2keV}$ by moving count rates in the soft and hard bands in turn both upward and downward by the error and repeating the process above. Additionally, given the relatively high \mbox{X-ray} counts of our quasars (the median number of counts in the 0.5--7~keV band is 47 with a range of \mbox{16--67}), we could also use XSPEC to fit the spectra with the same model to determine $\Gamma_{\rm eff}$, and the fitting results agree well with the estimations from {\tt\string modelflux}. Thus, we can use the XSPEC results for our quasars, which are more accurate, to compare with the {\tt\string modelflux} results for the T20 sample. The \mbox{X-ray} properties are reported in Table \ref{tab2}.

\subsection{Fitting optical spectra} \label{subsec:optical}
   We fitted all available SDSS spectra of the large \ion{C}{4} EW quasars, which are shown in Figure~\ref{fig1}b, including the two from the T20 quasar sample. We measured the \ion{C}{4} EW using the method outlined in T20, so that we can consistently compare the \ion{C}{4} line properties of our quasars with the large quasar sample in T20. We employed the PyQSOFit$\footnote{\url{https://github.com/legolason/PyQSOFit}}$  (\citealt{Guo2018}) software, which is based on the code used in \citet{Shen2011}. We first fitted the global continuum using a combination of a simple power-law, a low-order polynomial, and UV \ion{Fe}{2} templates. The line-free regions (at rest-frame wavelengths) are built into PyQSOFit. Next we measured the properties of the \ion{C}{4} emission line. We fit a local power-law continuum to the rest-frame relatively line-free regions of 1445--1465~\AA\ and 1690--1705~\AA. The fitted continuum was then subtracted from the \ion{C}{4} line region, and three broad Gaussian profiles were used to fit the emission line within \mbox{1500--1600~\AA}\ (see Figure~\ref{fig1}a). We masked the 3$\sigma$ outliers from the spectrum smoothed by a 20-pixel boxcar filter to reduce the impact of narrow absorption lines around the \ion{C}{4} line. PyQSOFit returned the EW based on the multi-Gaussian model, FWHM, peak wavelength, and other measurements of the \ion{C}{4} line.

Another high-ionization emission line, \ion{He}{2}$\ \lambda1640$, has also been used as a proxy for the amount of ionizing radiation in previous work. Our method of measuring the \ion{He}{2} EW is the same as in \citet{Timlin2021} (hereafter T21), also aiming to compare our quasars with the quasar sample with \ion{He}{2} measurements in T21. The local continuum in the \ion{He}{2} emission-line region of each spectrum was determined by fitting a linear model to the median values in the continuum windows \mbox{1420--1460~\AA}\ and \mbox{1680--1700~\AA}, and the 3$\sigma$ clipping method was also incorporated to remove the effects of any narrow spikes in these regions. The \ion{He}{2} EW was then measured by directly integrating the continuum-normalized flux in the window \mbox{1620--1650~\AA}. This method measures the emission where \ion{He}{2} dominates the spectrum. The \ion{He}{2} lines in all of our quasar spectra are well detected. The measurement results for the \ion{C}{4} and \ion{He}{2} lines are reported in Table \ref{tab3}.  \\

\subsection{Long-term light curves} \label{subsec:lc}

\begin{figure}[!t]
\centering
\includegraphics[width=0.5\textwidth]{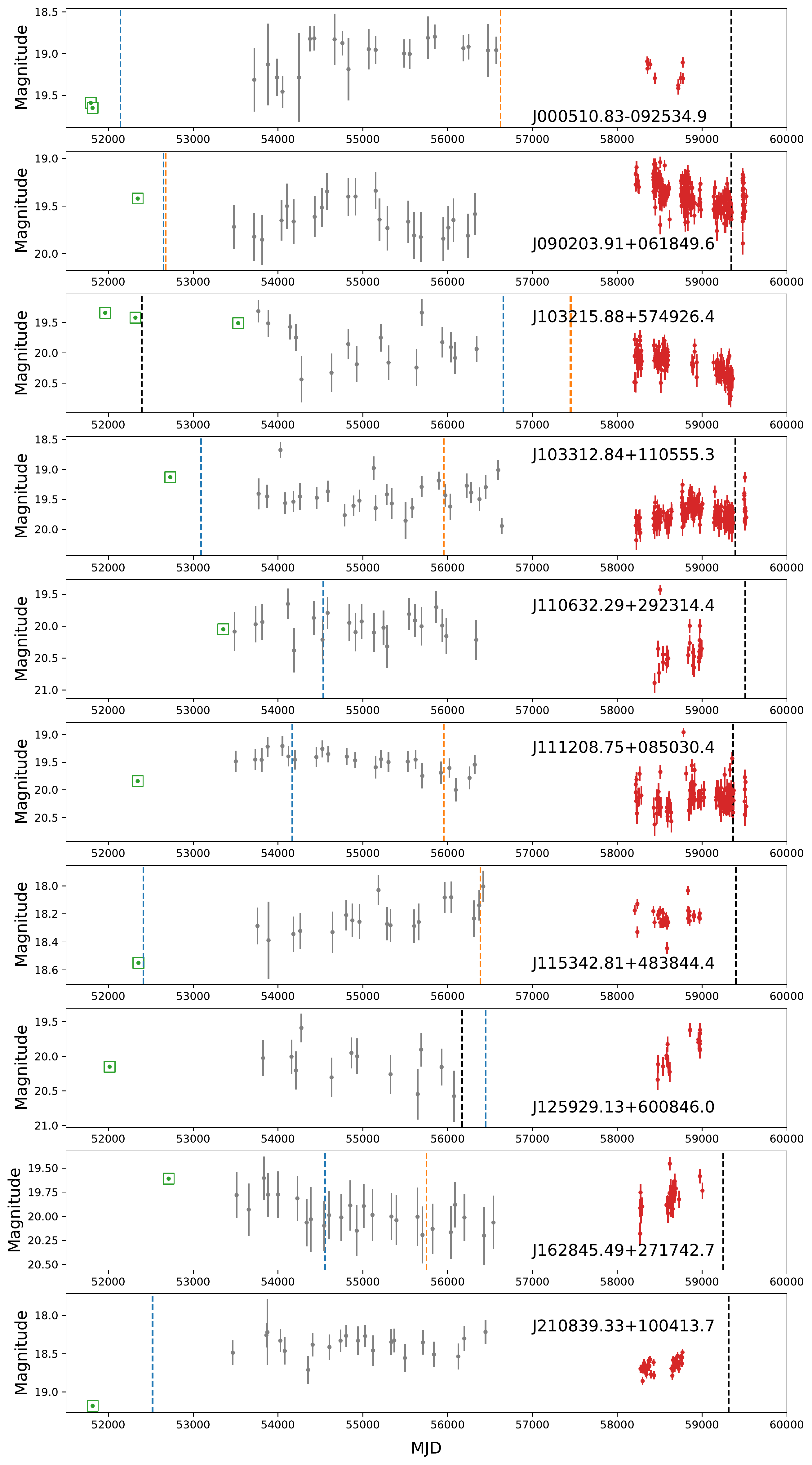}
\caption{The long-term light curves and timelines of multi-wavelength observations. The light curves include SDSS (green points), CRTS (grey points) and ZTF (red points) photometric data, all cross-calibrated to SDSS $g$-band magnitude, and we marked the dates of SDSS spectroscopic observations (blue dashed lines for DR7 spectra, and orange for the others) and the Chandra observations (black dashed lines). The CRTS light curves are binned to reduce noise. To correct for different filter curves, we convolved the SDSS spectra with the ZTF/CRTS/SDSS filter curves to obtain the magnitude corrections.  \label{figlc}}
\end{figure}

Because the \mbox{X-ray} observations, SDSS spectroscopic observations, and SDSS photometric observations are not simultaneous, we should assess the possible effects of quasar variability when we conduct multi-wavelength data analyses.
We generate long-term light curves to check if there are substantial changes in the luminosities of these quasars at the times of the SDSS spectroscopic and \mbox{X-ray} observations.

The light curves are shown in Figure~\ref{figlc}, including SDSS, the Catalina Real-time Transient Survey (CRTS, \citealt{Drake2009}), and the Zwicky Transient Facility (ZTF, \citealt{Bellm2019}) photometric data, and the dates of SDSS spectroscopic observations and the Chandra observations are marked. To correct for different filter curves, we convolved the nearest SDSS spectra with the ZTF/CRTS/SDSS filter curves to obtain the magnitude corrections, and all optical data are then cross-calibrated to SDSS $g$-band (\citealt{Yangqian2020}). While the calibration is slight and reliable for ZTF data since its filter curves are similar to SDSS,  the calibration for CRTS data is more difficult and may have large uncertainties because CRTS data are observed through a wide band with a resolving power of $\sim 1$ and thus variations of spectral shape can make substantial differences. 

It can be seen from the light curves that the magnitudes of the quasars have not increased or decreased substantially around the dates of the SDSS spectroscopic and \mbox{X-ray} observations. In addition, it has been suggested that single-epoch spectroscopic results do not significantly change where quasars are located in \ion{C}{4} parameter space on rest-frame timescales of $\approx 300$ days (\citealt{Rivera2020}). Therefore, we believe that quasar variability should not have material impacts upon the subsequent analyses.

In the multi-wavelength data analysis throughout the work, if not specified, we used the SDSS spectra and SDSS/ZTF photometric data closest to the \mbox{X-ray} observation dates to ensure as much consistency as possible between different bands. 

\section{BASIC OPTICAL/UV AND \mbox{X-ray} OBSERVATIONAL RESULTS} \label{sec:prop}

\begin{figure*}[!t]
\centering
\includegraphics[width=\textwidth]{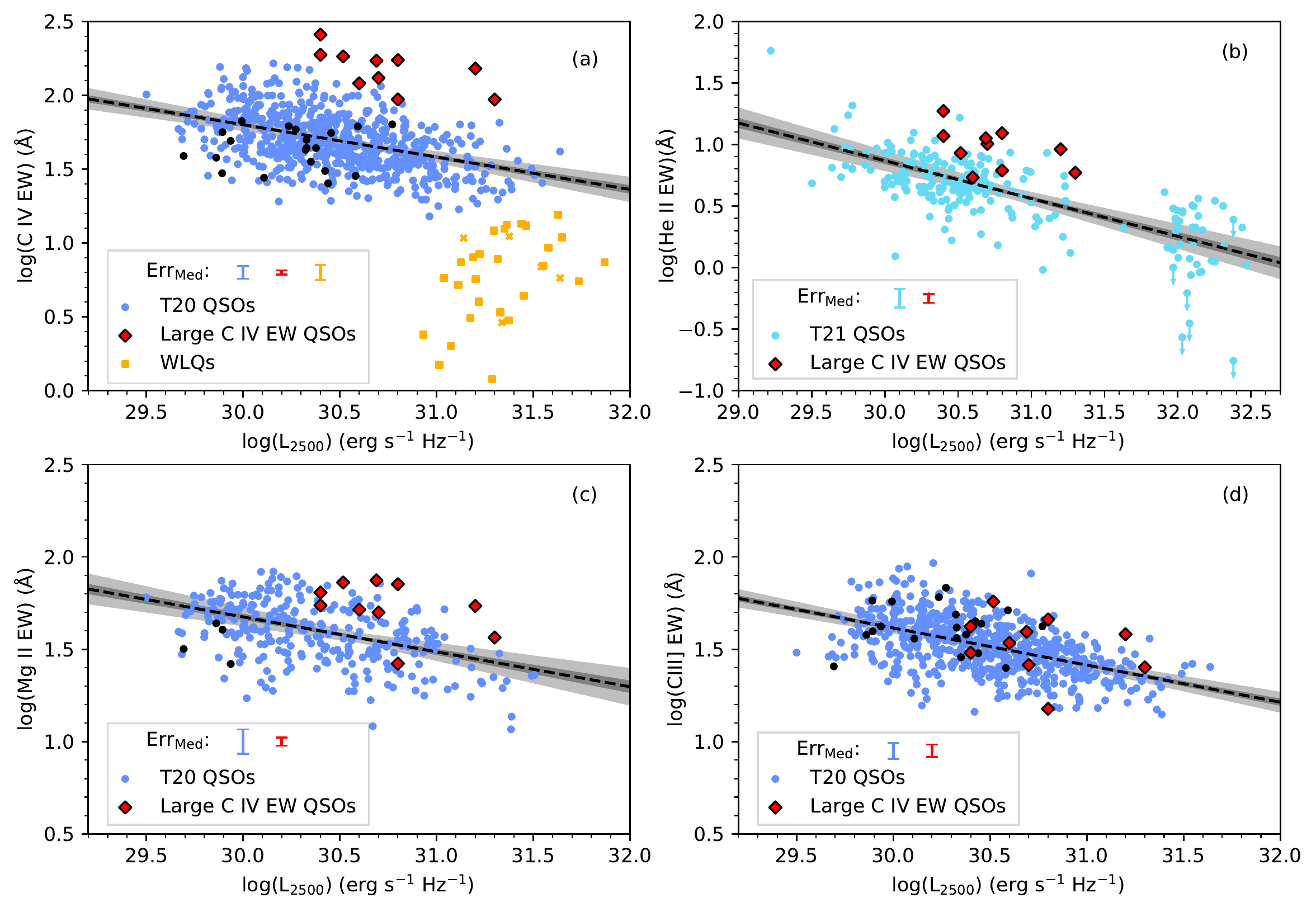}
\caption{(a) The Baldwin effect (dashed black line; grey/dark grey shaded region depicts the 3$\sigma$/1$\sigma$ confidence interval) derived using the 637 typical quasars in the T20 \ion{C}{4} subsample (blue points; black points represent quasars with \mbox{X-ray} upper limits) and the large \ion{C}{4} EW quasars (red diamonds). WLQs (orange squares; orange crosses indicate WLQs with \mbox{X-ray} upper limits) from the \citet{Ni2018,Ni2022} Representative sample are also included for comparison. The median errors on log(\ion{C}{4} EW) of the three samples are represented by error bars in the corresponding colors (labeled as ``$\rm Err_{Med}$''), and the measurement error of $L_{\rm 2500}$ is too small to be depicted. Clearly the WLQs and the large \ion{C}{4} EW quasars are extreme outliers from the general population in T20.  
(b) The Baldwin effect between log(\ion{He}{2} EW) and log($L_{\rm 2500}$) (dashed black line; grey/dark grey shaded region depicts the 3$\sigma$/1$\sigma$ confidence interval); this relation is used to compute $\Delta$log(\ion{He}{2} EW). The light blue points are 206 quasars from the T21 \ion{He}{2} sample, the downward-pointing arrows depict upper limits on the \ion{He}{2} EW when the emission line is not detected, and the red diamonds are our large \ion{C}{4} EW quasars; their median errors on log(\ion{He}{2} EW) are indicated by the error bars. Larger \ion{He}{2} EWs of the large \ion{C}{4} EW quasars also indicate a larger number of ionizing photons reaching the high-ionization broad emission-line region.    
(c,d) The same as panel (a), replacing \ion{C}{4} EW with \ion{Mg}{2} EW and \ion{C}{3]} EW.  \label{fig2}}
\end{figure*}

\subsection{Large emission-line EWs} \label{subsec:largeEW}

It is well known that the emission-line EWs of quasars are related to their luminosities (the Baldwin effect; \citealt{Baldwin1977}), i.e., the emission-line EW decreases with increasing continuum luminosity. Under our source-selection criteria in Section~\ref{subsec:sample}, the \ion{C}{4} EWs of our eight quasars generally are significantly larger than those of other quasars in T20 with similar UV luminosities (see Figure~\ref{fig2}a). As mentioned in Section~\ref{subsec:sample}, only two quasars in T20 have similar $L_{2500}$ values and \ion{C}{4} EWs, and their \mbox{X-ray} and optical properties are also similar to those of our new sample, which will be shown in the results below, and thus we included them in the following analyses. We fitted a linear relation between log(\ion{C}{4} EW) and log($L_{\rm 2500}$) to the T20 \ion{C}{4} subsample of 637 typical quasars, together with our large \ion{C}{4} EW quasars.

We also compared the \ion{He}{2} EWs of our quasars with the large \ion{He}{2} EW sample in T21. The \ion{He}{2} EWs of the T21 quasar sample are strongly related to $L_{\rm 2500}$, and our large \ion{C}{4} EW quasars also have large \ion{He}{2} EWs (see Figure~\ref{fig2}b), which further indicates a larger number of ionizing photons reaching the high-ionization broad emission-line region. For the \mbox{log(\ion{He}{2} EW)--log($L_{\rm 2500}$)} relation, we also used a linear relation to fit the 206-quasar sample from T21. Note that some of the \ion{He}{2} EWs in T21 are upper limits, and thus throughout this work we used the Bayesian fitting method developed in \citet{Kelly2007} and implemented in the linmix\footnote{\url{https://linmix.readthedocs.io/en/latest/maths.html}} Python package, which can incorporate measurements that are upper (or lower) limits, including errors in both dimensions. 

Even though our large \ion{C}{4} EW quasars improve the coverage in the above two parameter spaces, they are numerically overwhelmed by the 637 typical quasars in the T20 \ion{C}{4} subsample or the 206 in the T21 sample, so that the best-fit linear models\footnote{log$_{10}$(\ion{C}{4} EW)=($-$0.220 $\pm$ 0.018)log$_{10}$($L_{2500}$)+(8.392 $\pm$ 0.555) and log$_{10}$(\ion{He}{2} EW)=($-$0.308 $\pm$ 0.022)log$_{10}$($L_{2500}$)+(10.094 $\pm$ 0.684)} are almost identical to those obtained in T20 and T21 (see T20 Equation 8 and T21 Table~2), and other relations we derive below are in the same situation. Thus, the linear relations provide general trends for typical quasars that can be compared with our sample.

We also considered \ion{Mg}{2} and \ion{C}{3]}, two low-ionization emission lines visible in the spectra, and compared their EWs with those of the T20 sample. Compared with the T20 sample, the \ion{Mg}{2} EWs of our quasars are generally larger (see Figure~\ref{fig2}c), while the \ion{C}{3]} EWs show no significant difference (see Figure~\ref{fig2}d). As mentioned in Section~\ref{subsec:lc}, except for the two quasars from the T20 sample, for the other eight quasars we use the latest SDSS spectra and ZTF photometric data in the analysis, namely the observation results closest in time to our X-ray observations.

\begin{figure}[!t]
\centering
\includegraphics[width=0.5\textwidth]{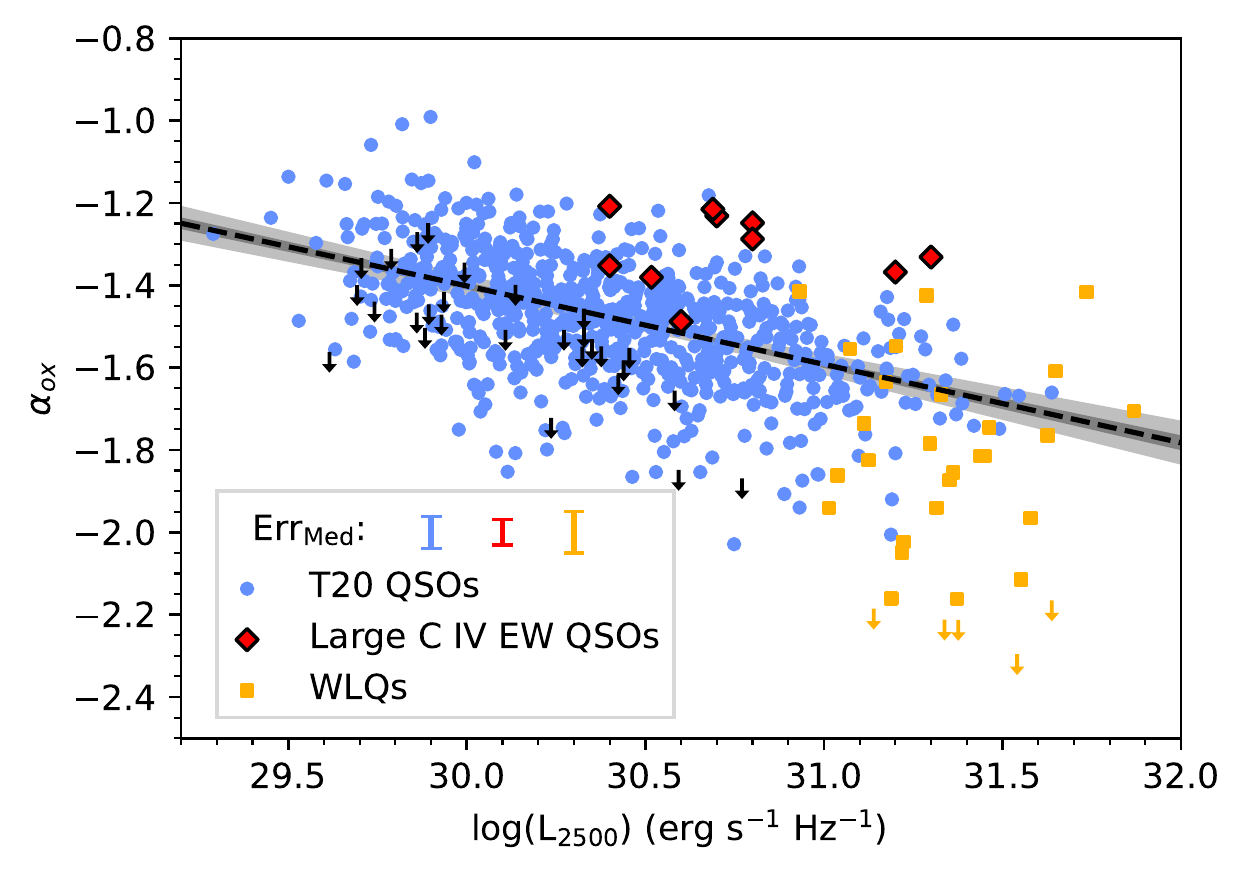}
\caption{ Dependence of $\alpha_{\rm ox}$ on the 2500~\AA\ monochromatic luminosity, $L_{\rm 2500}$. Blue points represent the ``Sensitive'' sample of 753 typical quasars in T20 and black arrows depict quasars with \mbox{X-ray} upper limits, red diamonds are our large \ion{C}{4} EW quasars, and orange squares and orange arrows represent WLQs and WLQs with \mbox{X-ray} upper limits; their median errors on $\alpha_{\rm ox}$ are shown by error bars in the corresponding colors. The black dashed line depicts the best-fit relation between $\alpha_{\rm ox}$ and log($L_{\rm 2500}$) and is used to compute $\Delta\alpha_{\rm ox}$; the 3$\sigma$/1$\sigma$  confidence interval is shown as the grey/dark grey shaded region. The WLQs are often \mbox{X-ray} weaker than what the $\alpha_{\rm ox}-L_{\rm 2500}$ relation predicts while our large \ion{C}{4} EW quasars are generally \mbox{X-ray} stronger. \label{figXexcess}}
\end{figure}

\begin{figure}[!t]
\centering
\includegraphics[width=0.5\textwidth]{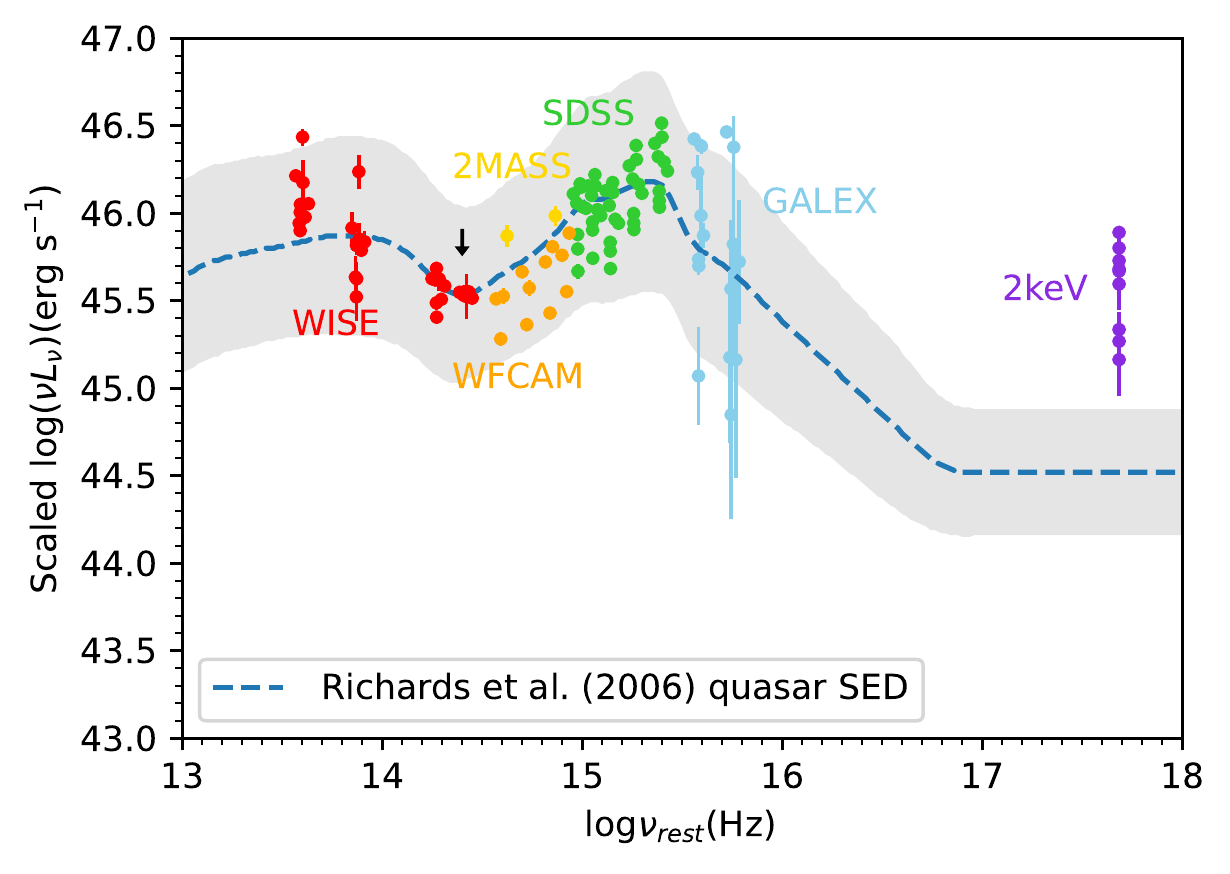}
\caption{Combined SED for the large \ion{C}{4} EW quasars. The SED for each object was scaled to the composite quasar SED of optically luminous quasars (dashed line, and $1\sigma$ error shown with grey region; \citealt{Richards2006}) at rest-frame $10^{14.4}$ Hz (black arrow), and then combined. The combined IR-to-UV SED of the large \ion{C}{4} EW quasars is similar to the composite quasar SED, except for the apparent \mbox{X-ray} excess.  \label{figSED}}
\end{figure}

\subsection{\mbox{X-ray} excess} \label{subsec:Xexcess}

In addition to emission-line EWs, another commonly used indirect measurement of the strength of the ionizing emission in quasars is the optical/UV-to-\mbox{X-ray} power-law spectral slope, $\alpha_{\rm ox}$. We also plausibly expected our large \ion{C}{4} EW quasars to have larger $\alpha_{\rm ox}$ than the predictions from the standard \mbox{$\alpha_{\rm ox}$--$L_{\rm 2500}$} relation ($\alpha_{\rm ox}$ values are reported in Table \ref{tab2}.). The result was as expected (see Figure~\ref{figXexcess}). We also depicted 32 WLQs from the \citet{Ni2018,Ni2022} ``Representative" sample for comparison. It is notable that the WLQs are often \mbox{X-ray} weaker than what the \mbox{$\alpha_{\rm ox}$--$ \rm 2500\ \AA$} relation predicts, while our large \ion{C}{4} EW quasars are generally \mbox{X-ray} stronger.

The \mbox{X-ray} excess can also be seen in the combined spectral energy distribution (SED) (see Figure~\ref{figSED}). The IR-to-UV SED photometric data were from the \citet{Paris2018} catalog. These data have been corrected for Galactic extinction following the dereddening approach of \citet{Cardelli1989} and \citet{Donnell1994} and have been corrected for intergalactic medium extinction following \citet{Meiksin2006}. We compared this composite SED with the composite quasar SED of optically luminous quasars from \citet{Richards2006}. The combined IR-to-UV SED of the large C IV EW quasars is similar to the composite quasar SED, but the \mbox{X-ray} emission shows an apparent excess. Note that the apparent SDSS $u$-band and $g$-band deviations are caused by the strong Ly$\alpha$ and \ion{C}{4} emission lines, which are estimated from the spectra to cause an excess of about 0.2 dex.
                                                                                                                                                                                                                                                                                                                                                                                                                                                                                                                  
\begin{figure*}[!t]
\gridline{\fig{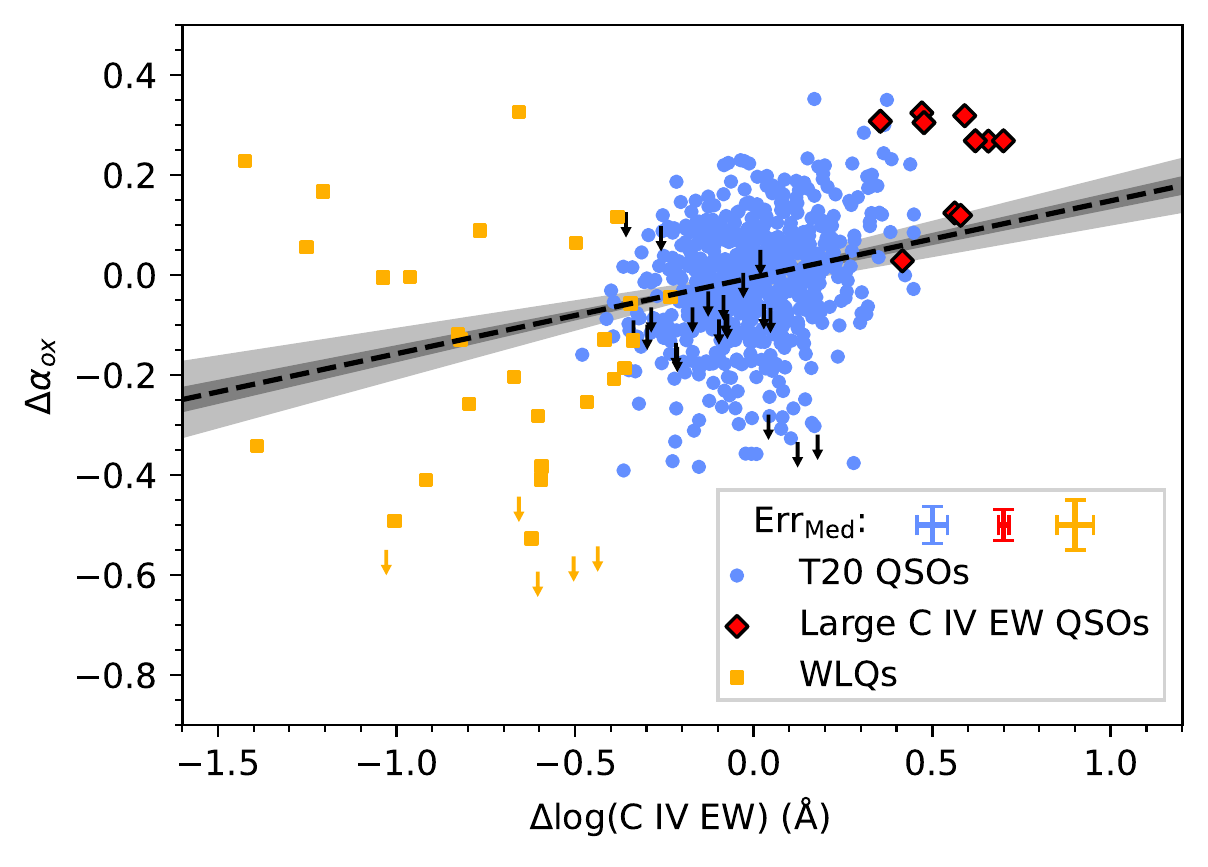}{0.5\textwidth}{(a)}
          \fig{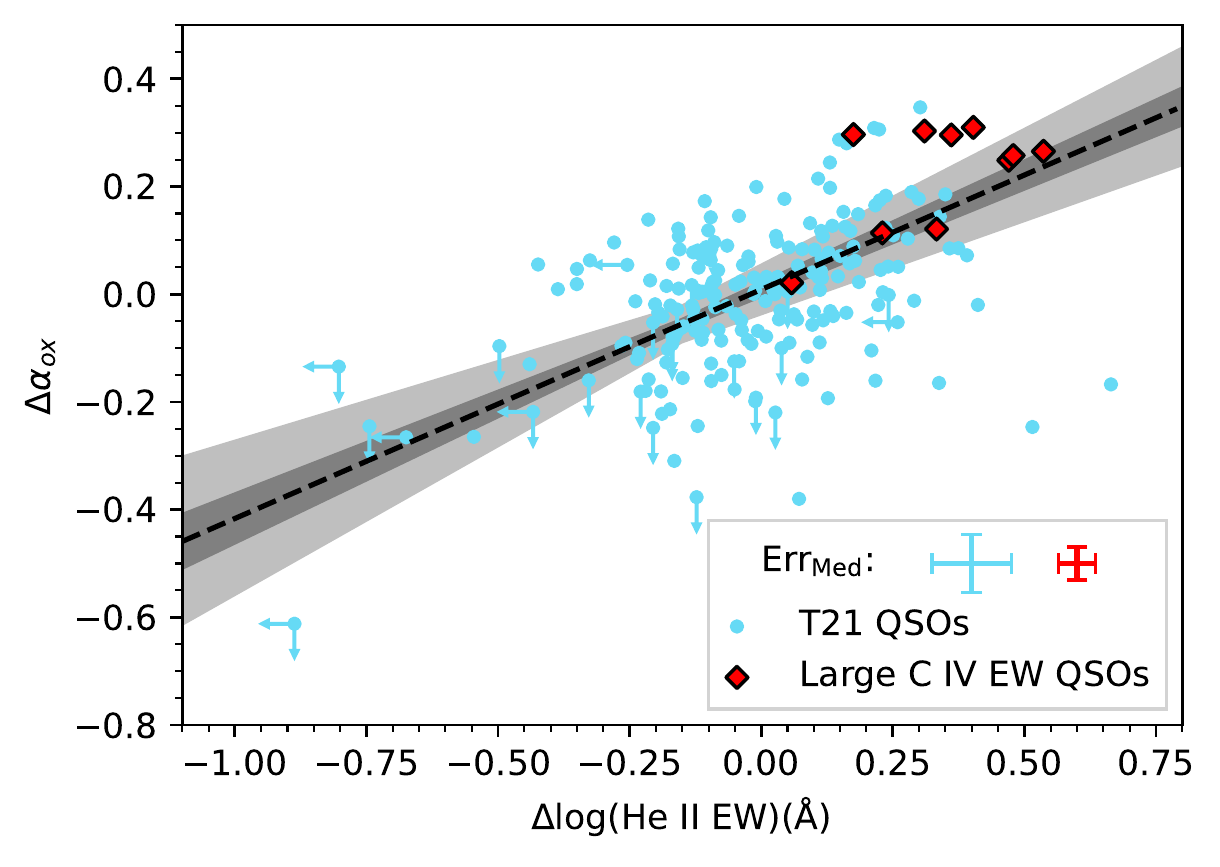}{0.5\textwidth}{(b)}}
\caption{(a) The luminosity-adjusted $\alpha_{\rm ox}$, $\Delta\alpha_{\rm ox}$, as a function of luminosity-adjusted \ion{C}{4} EW, $\Delta$log(\ion{C}{4} EW); the color scheme is the same as in Figure~\ref{fig2}, and the median error on $\Delta\alpha_{\rm ox}$ and $\Delta$log(\ion{C}{4} EW) of each sample is shown by the error bar in the corresponding colors. The general \mbox{X-ray} excess of our large \ion{C}{4} EW quasars is also apparent with respect to their $\Delta$log(\ion{C}{4} EW), in broad accord with the scenario that extremely strong ionizing radiation leads to larger \ion{C}{4} EW in these quasars. 
(b) $\Delta\alpha_{\rm ox}$ as a function of luminosity-adjusted \ion{He}{2} EW, $\Delta$log(\ion{He}{2} EW). The light blue points are T21 quasars; left-pointing arrows depict upper limits of the $\Delta$(\ion{He}{2} EW), and downward-pointing arrows depict upper limits of $\Delta\alpha_{\rm ox}$ for \mbox{X-ray} non-detections. The red diamonds are our large \ion{C}{4} EW quasars. Error bars of $\Delta\alpha_{\rm ox}$ and $\Delta$log(\ion{He}{2} EW) are shown in the bottom-right corner.  \label{fig4}}
\end{figure*}

Using the relations between $\alpha_{\rm ox}$, \ion{C}{4} EW, \ion{He}{2} EW, and $L_{\rm 2500}$, respectively, we can derive the luminosity-adjusted values, $\Delta\alpha_{\rm ox}$, $\Delta$log(\ion{C}{4} EW)\footnote{Like $\Delta\alpha_{\rm ox}$, $\Delta$log(\ion{C}{4} EW) is the luminosity-adjusted value defined as observed log(\ion{C}{4} EW) minus the prediction from the \mbox{log(\ion{C}{4} EW)--$L_{\rm 2500}$} relation in T20; $\Delta$log(\ion{He}{2} EW) is defined in the same way.}, and $\Delta$log(\ion{He}{2} EW). The $\Delta\alpha_{\rm ox}$ values of our large \ion{C}{4} EW quasars are generally larger than the linear \mbox{$\Delta$log(\ion{C}{4} EW)--$\Delta\alpha_{\rm ox}$} relation prediction and perhaps also the \mbox{$\Delta$log(\ion{He}{2} EW)--$\Delta\alpha_{\rm ox}$} relation prediction (see Figure~\ref{fig4}). The \mbox{X-ray} excess of the large \ion{C}{4} EW quasars in the \mbox{$\Delta$log(\ion{He}{2} EW)--$\Delta\alpha_{\rm ox}$} space is not as large as in the $\Delta$log(\ion{C}{4} EW)$-$$\Delta\alpha_{\rm ox}$ space, which is partly because of higher uncertainty in the \ion{He}{2} EW measurements and thus a larger uncertainty of the derived correlation.

\subsection{The \ion{C}{4} EW -- \ion{C}{4} blueshift relation} \label{subsec:EW_blsft}

\begin{figure}[!t]
\centering
\includegraphics[width=0.5\textwidth]{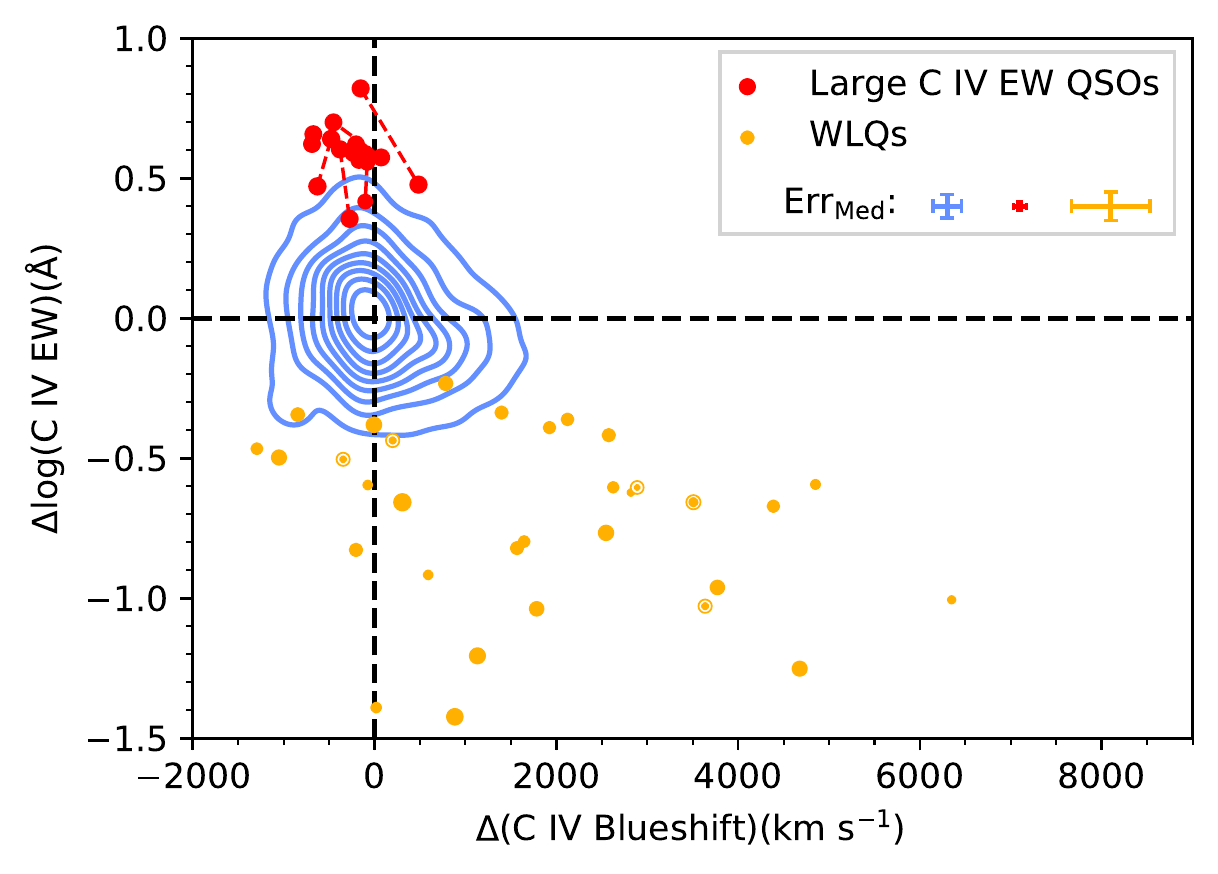}
\caption{The $\Delta$log(\ion{C}{4} EW) with respect to $\Delta$(\ion{C}{4} blueshift); both are luminosity-adjusted values. The blue contours depict the distribution of T20 quasars. Red points are our large \ion{C}{4} EW quasars, orange points represent WLQs, and WLQs with X-ray upper limits are circled; the sizes of the points indicate the $\Delta\alpha_{\rm ox}$ values. All spectral observation results of the large \ion{C}{4} EW quasars are shown, and points of the same source are connected by dashed lines. The vertical and horizontal dashed lines represent the expected \ion{C}{4} blueshift and \ion{C}{4} EW according to $L_{\rm 2500}$. Median errors for T20 quasars, the large \ion{C}{4} EW quasars, and WLQs are shown by the error bars in the corresponding colors. Clearly our large \ion{C}{4} EW QSO sample has strong \mbox{X-ray} emission and small blueshift, representing the ``opposite" extreme from the WLQs in this parameter space. \label{fig5}}
\end{figure}

We compared the large \ion{C}{4} EW quasars with T20 quasars and WLQs in the \mbox{$\Delta$log(\ion{C}{4} EW) -- $\Delta$(\ion{C}{4} blueshift)}\footnote{We define the \ion{C}{4} blueshift as $c(1549.06 - \lambda_{\rm peak})/1549.06$, measured in units of $\rm km\ s^{-1}$, where $\lambda_{\rm peak}$ is the measured peak of the emission line (in $\rm \AA$), 1549.06 $\rm \AA$ is the laboratory wavelength of the \ion{C}{4} emission line (see Table 4 of \citealt{Vanden2001}), and $c$ is the speed of light. See Table \ref{tab3} for measured \ion{C}{4} blueshift values for all ten quasars.} space (see Figure~\ref{fig5}). For quasars that have more than one SDSS spectral observation, we show the measurements of all spectra to inspect the impact of \ion{C}{4} variability on the position, and points of the same source are connected by dashed lines. The sizes of the points indicate the $\Delta\alpha_{\rm ox}$ values. Our quasars (diamonds) occupy the upper-left corner of the space, meaning that they generally have larger \ion{C}{4} EW, smaller blueshifts ($\Delta$(\ion{C}{4} blueshift) $< 0$), and stronger \mbox{X-ray} emission compared to the T20 typical quasar sample. The \ion{C}{4} line variability does not change this trend (also see \citealt{Rivera2020}). 

WLQs usually have small \ion{C}{4} EWs, large \ion{C}{4} blueshifts, and weak \mbox{X-ray} emission. In this sense, our large \ion{C}{4} EW quasars seem to be ``anti-WLQs", representing the ``opposite'' extreme from WLQs in Figure~\ref{fig5}. Many properties of WLQs can be explained by a ``shielding” model, in which a geometrically and optically thick inner accretion disk and its associated wind, expected for a quasar accreting at a high Eddington ratio, both prevents ionizing EUV/\mbox{X-ray} photons from reaching the high-ionization broad emission-line region and also sometimes blocks the line-of-sight to the central \mbox{X-ray} emitting region (e.g., \citealt{Ni2018,Ni2022}). Thus, WLQs have weak emission lines and often weak \mbox{X-ray} emission with signs of heavy intrinsic \mbox{X-ray} absorption. If the large \ion{C}{4} EW quasars represent the ``opposite" extreme from WLQs, they might be expected to have relatively low Eddington ratios and little intrinsic absorption.

\subsection{Assessing causes of the strong X-ray emission} \label{subsec:Why_excess}

\begin{figure*}[!t]
\gridline{\fig{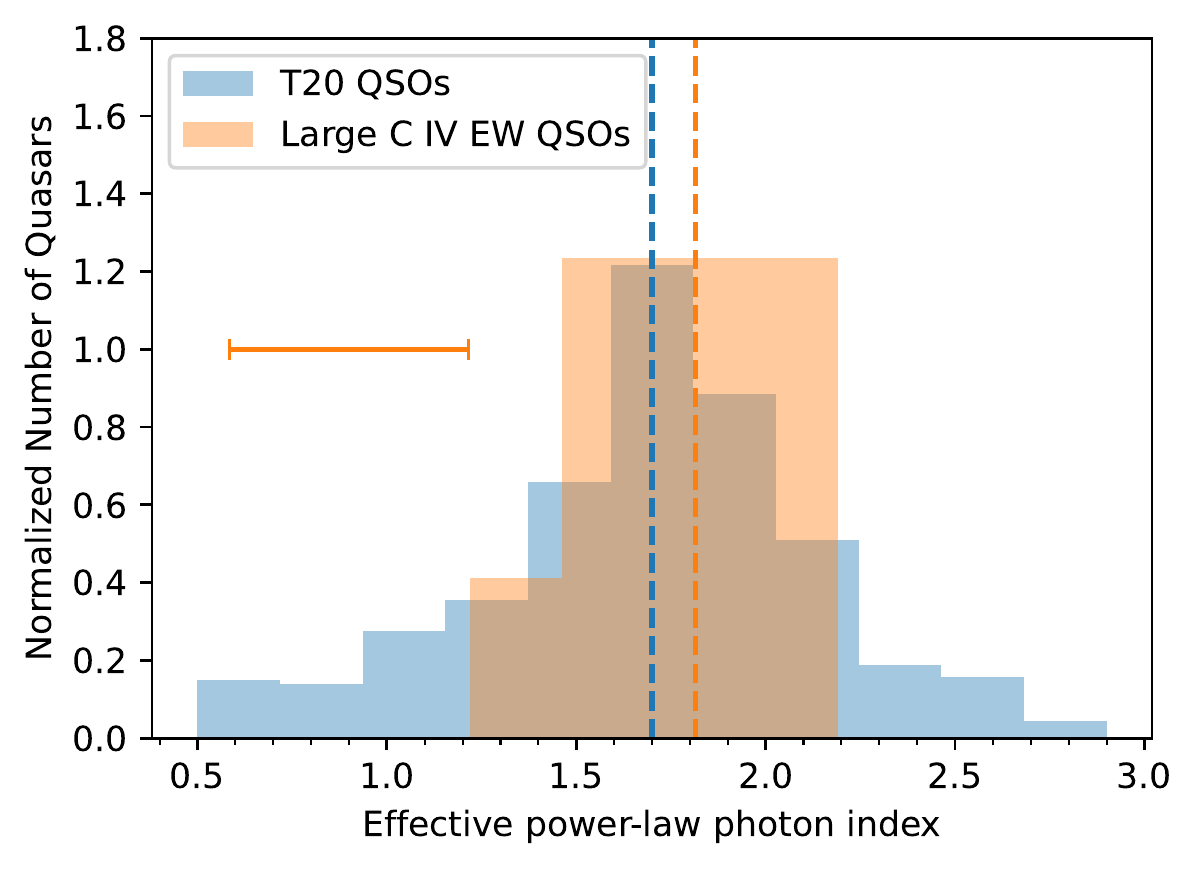}{0.5\textwidth}{(a)}
          \fig{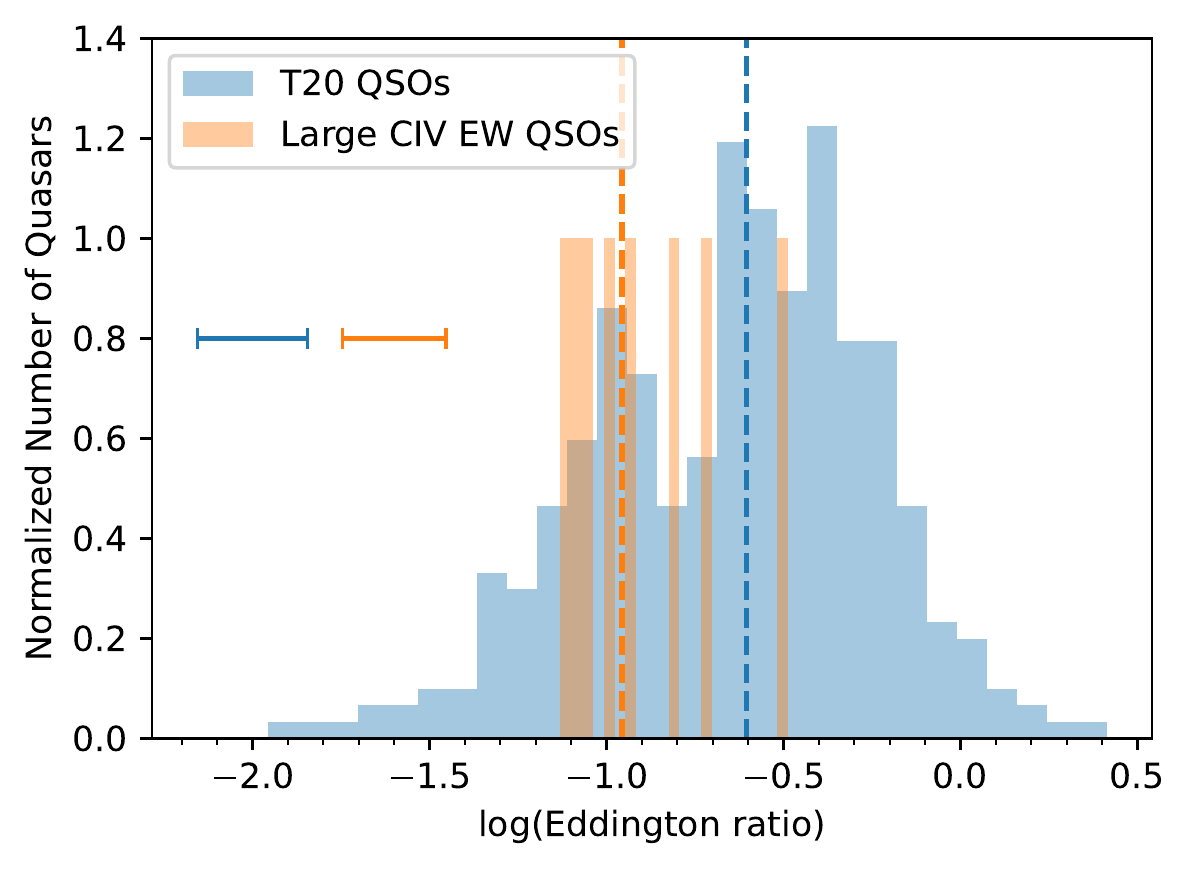}{0.5\textwidth}{(b)}}
\caption{(a) The distribution of effective power-law photon indexes ($\Gamma_{\rm eff}$) for T20 quasars and our large \ion{C}{4} EW quasars. The median $\Gamma_{\rm eff}$ values are shown by the vertical dashed lines in corresponding colors. The median error on $\Gamma_{\rm eff}$ of large \ion{C}{4} EW quasar sample, estimated from XSPEC, is shown with the error bar. The $\Gamma_{\rm eff}$ of our large \ion{C}{4} EW quasars do not deviate apparently from T20 typical quasars. 
(b) The distribution of log(Eddington ratio) for T20 quasars and our large \ion{C}{4} EW quasars. The median log(Eddington ratio) values are shown by the vertical dashed lines in corresponding colors. The median errors on log(Eddington ratio) for both samples are shown with the error bars; note that these errors only included measurement uncertainties but not the statistical uncertainty from virial mass calibrations or the systematic uncertainties for the virial black-hole masses (\citealt{Shen2011}). \label{fig6}}
\end{figure*}

To attempt to assess what causes the strong \mbox{X-ray} emission of the large \ion{C}{4} EW quasars, we tried to find clues from their \mbox{X-ray} spectra and Eddington ratios.
We fitted the \mbox{X-ray} spectra using XSPEC again, adding an intrinsic absorption component to the model. The resulting hydrogen column densities are small ($N\rm _H < 10^{22} \ cm^{-2}$), just as we expected, so that intrinsic absorption effects are small, and the effective power-law photon indices ($\Gamma_{\rm eff}$) provide an appropriate measure of the intrinsic spectral shape. No other distinctive features can be seen in the spectra, partly because of the limited counts, so that few clues about the \mbox{X-ray} excess can be obtained via \mbox{X-ray} spectral fitting. The median $\Gamma_{\rm eff}$ of the large \ion{C}{4} EW quasars is $1.81\pm0.13$ (the $1\sigma$ uncertainty is derived from bootstrap resampling), consistent with the median value of 1.9 for typical quasars (e.g., \citealt{Scott2011}). We also compared the distribution of their $\Gamma_{\rm eff}$ with those of the T20 quasars (Figure~\ref{fig6}a), but they do not show a significant difference. We performed a two-sample Anderson-Darling test to quantify the difference, if any, between the two distributions, and the null-hypothesis probability is $P_{\rm null}=0.55$, meaning the $\Gamma_{\rm eff}$ values of our large \ion{C}{4} EW quasars are not demonstrably different from those of the T20 quasars. While any solid conclusions about the $\Gamma_{\rm eff}$ values of our targeted quasars must await higher quality \mbox{X-ray} spectra, we do note that our median $\Gamma_{\rm eff}$  of $1.81\pm0.13$ is consistent with fairly low Eddington ratios (e.g., \citealt{Shemmer2008,Brightman2013}).

It has been suggested that the 2--10 keV bolometric corrections ($L_ {\rm bol}/L_{\rm X}$) of quasars are positively correlated with their Eddington ratios (e.g., \citealt{Vasudevan2007,Lusso2012}), and our large \ion{C}{4} EW quasars have low $L_ {\rm bol}/L_{\rm X}$ which might imply low Eddington ratios. Besides, if indeed WLQs have high Eddington ratios, then our apparent ``anti-WLQs" may also have low Eddington ratios in this sense. Thus, we estimated the Eddington ratios of our large \ion{C}{4} EW quasars. We obtain bolometric luminosities ($L_ {\rm bol}$) of our targets via integrating the SED from 1~micron to 10~keV, and the mutiwavelength photometric data come from the \citet{Paris2018} catalog (see Figure~\ref{figSED}). The black-hole masses are from the \citet{Shen2011} catalog. All of our quasars have both \ion{C}{4} and \ion{Mg}{2} in their spectra, and we only use black-hole masses estimated from \ion{Mg}{2} FWHMs to avoid systematic discrepancies caused by different ``single-epoch” black-hole mass estimation methods (e.g., \citealt{Shen2008}). The black-hole mass estimator is consistent in subsequent works (e.g., \citealt{Shen2012,Trakhtenbrot2012}). The Eddington ratios are listed in Table \ref{tab1} and shown in Figure~\ref{fig6}b. We also plotted the distribution of Eddington ratios of the T20 quasars which have black-hole masses estimated from \ion{Mg}{2} FWHMs (we directly use $L_ {\rm bol}$ from the \citealt{Shen2011} catalog for the T20 quasars, which are appropriate in an average sense). The Eddington ratios of our sample, with a median value of 0.11, are suggestively smaller than those of T20 quasars according to an Anderson-Darling test ($P_{\rm null}=0.007$). However, we note there are large uncertainties in Eddington-ratio estimation, especially in the black-hole mass measurements, and thus a larger sample is needed to draw a conclusion.

\section{DISCUSSION} \label{sec:disc}

Above we have shown that luminous quasars selected to have the most extreme
\ion{C}{4} EW  are characterized by excess EW in \ion{He}{2} and \ion{Mg}{2},
but only marginal excess \ion{C}{3]} EW (see Figure~\ref{fig2}). In addition, there 
is clear excess \mbox{X-ray} emission (see Figures~\ref{figXexcess}--\ref{fig4}). What is the physical mechanism 
driving these unusual properties? Possible solutions are described below.                                                                                    

The EW of the broad emission lines depends, among other things, on the covering
factor of the broad line region gas. Are these high \ion{C}{4} EW AGN characterized 
by a very high covering factor? If so, then all lines should be
enhanced by about the same factor. As described above, the \ion{C}{3]} EW shows only 
a slight enhancement, if any, which clearly excludes this option. 

Another way to enhance the \ion{C}{4} EW, without changing the broad line region 
covering factor, is by lowering the observed local continuum luminosity, without changing
the continuum luminosity that illuminates the broad line region. If the optical-UV
continuum in quasars is produced by an accretion disk, the observed luminosity
is expected to be inclination dependent. Do high \ion{C}{4} EW quasars have a nearly
edge-on and limb-darkened view? In this case as well, all UV lines should be enhanced 
in about a similar way. So, as noted above, this possibility is also excluded.

We clearly need a mechanism that selectively enhances the \ion{C}{4} EW.
Another possible mechanism is an enhanced covering factor due to the addition 
of a low column density gas. If the gas ionization is matter bounded, rather than 
ionization bounded, it will produce only high-ionization lines. Such a component 
indeed characterizes the so-called ``wind component" of the broad line region gas,
characterized by a strong blueshift, and is seen clearly only in high-ionization lines,
in particular in \ion{C}{4}. Such a component can explain the lack of significant
lower ionization \ion{C}{3]} emission. However, such a low-column component will
not produce any \ion{Mg}{2} emission. The observed significant enhancement of
\ion{Mg}{2} emission clearly excludes this mechanism.

Since the excess \ion{C}{4} EW is associated with excess \mbox{X-ray} emission,
can the observed peculiar pattern of different UV line enhancements be explained
by an ionizing continuum effect?

\subsection{Comparison with photoionization calculation results}\label{subsec:link}

Below we show that the ionizing continuum shape can indeed explain a significant
part of the observed trends, but that metallicity also plays a significant role,
and its inclusion provides a nearly perfect explanation.
 
\subsubsection{The effect of the ionizing continuum}\label{subsubsec:ionizcont}

We use the photoionization code Cloudy \citep{Ferland1998}, in the mode which includes the
 radiation pressure compression (RPC) effect. This mode includes both the energy and 
 momentum transmitted to the gas by the incident photoionizing radiation which is absorbed. 
 As a result the gas density is not a free parameter, and is also not uniform. The density 
 structure within the photoionized slab is set by the absorbed radiation, and the resulting 
 ionization 
structure is independent of distance from the ionizing source. The remaining free
parameters are the spectral hardness of the ionizing continuum and the gas metallicity.
We use the RPC solutions for gas in the broad line region as reported by
\citet{Baskin2014}; particularly, see their Figure~5. 

Table~\ref{tabRPC}, in its first row, reports the maximal EWs of the \ion{C}{4}, \ion{He}{2}, 
\ion{Mg}{2}, and \ion{C}{3]} lines for a Solar metallicity ($Z=1$) gas and a typical ionizing 
continuum slope (from 1000~\AA \ to 1~keV) of $\alpha_{\rm ion}=-1.6$, which corresponds to 
$\alpha_{\rm ox}=-1.45$. The second row reports the effect on the EWs of a harder ionizing 
continuum of $\alpha_{\rm ion}=-1.2$, which corresponds to $\alpha_{\rm ox}=-1.16$. 
The two model values of $\alpha_{\rm ox}$ are comparable to the mean $\alpha_{\rm ox}$ 
values of the typical quasar sample and our large \ion{C}{4} EW quasar sample.

The last row in Table~\ref{tabRPC} gives the observed EW enhancement, 
$\Delta$EW/EW, for each of the lines, which is derived as follows. For each object we found 
$\Delta$EW, that is observed EW minus expected EW from the Baldwin relation (the best-fit 
dashed line in Figure~\ref{fig2}), and then calculate the mean fractional increase 
$\Delta$EW/expected EW. 
For the seven objects with more than one spectroscopic observation, the mean \ion{C}{4} EW
 obtained with the latest observation is $17\%$ smaller than that obtained with earlier 
 observations, so we corrected the \ion{C}{4} EW of the other three objects by a factor of 0.83 
 when calculating the mean fractional increase to account for this variability bias.

From rows 1 and 2, the \ion{C}{4} line EW increases by a factor of 2.6, which corresponds to 
$\Delta \rm EW/EW=1.6$, significantly lower than the observed 
$\Delta \rm EW/EW=2.33\pm 0.14$. For the
\ion{He}{2} EW the predicted and observed $\Delta \rm EW/EW$ match well, 1.31 versus
$1.23\pm 0.22$. For the \ion{Mg}{2} line the predicted $\Delta \rm EW/EW=1.35$ 
which is heavily offset from the observed value of $0.61\pm 0.10$. 
For \ion{C}{3]} (which is inevitably blended with \ion{Si}{3]}) the fit is marginal,
with predicted 0.49 versus $0.2\pm 0.1$.

\begin{figure}[!t]
\centering
\includegraphics[width=0.5\textwidth]{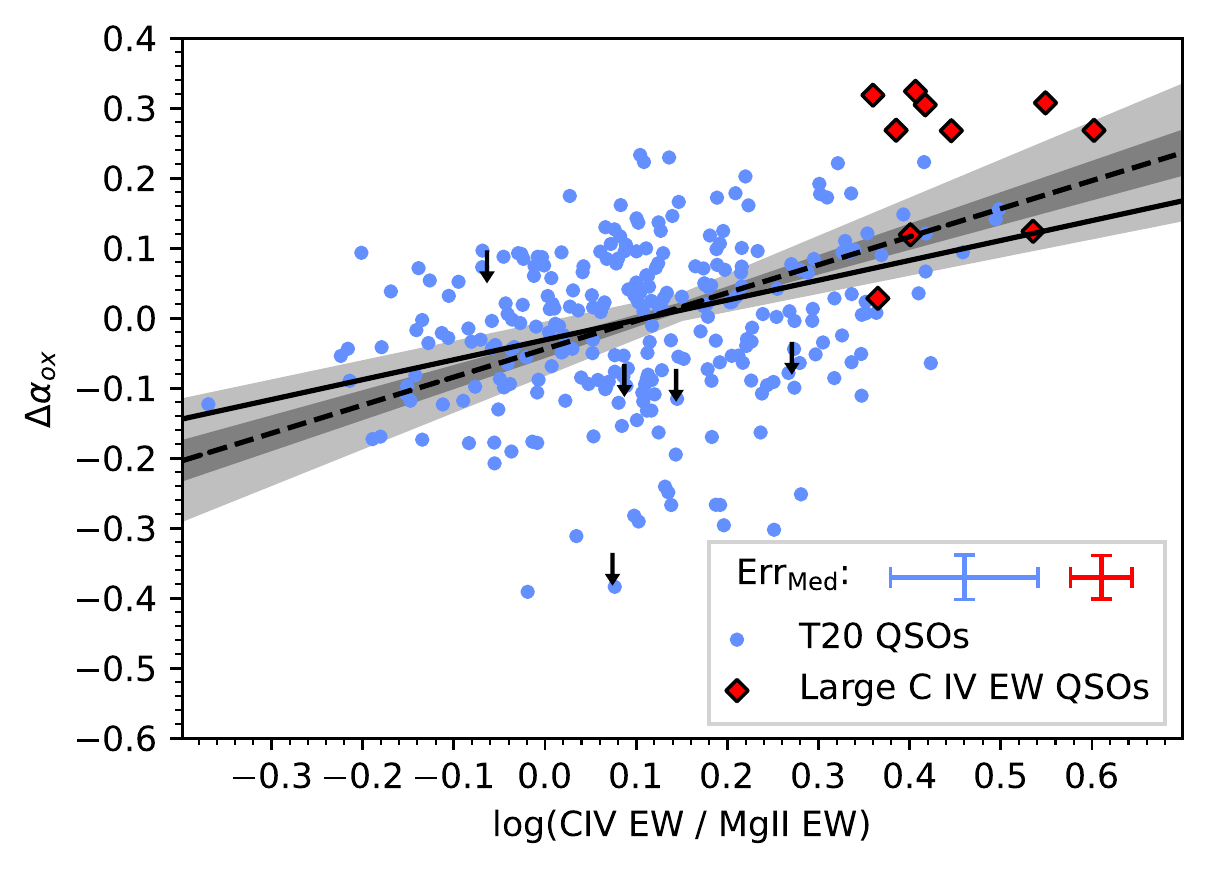}
\caption{$\Delta\alpha_{\rm ox}$ as a function of log(\ion{C}{4} EW / \ion{Mg}{2} EW). The color scheme is the same as in Figure~\ref{figXexcess}, and the median error on $\Delta\alpha_{\rm ox}$ and log(\ion{C}{4} EW / \ion{Mg}{2} EW) for each sample is shown by the error bars in the corresponding colors. The black solid line is the best-fit relation from T20. The dashed line and dark grey/grey regions are the linear relation and 1$\sigma$/3$\sigma$ confidence intervals obtained by fitting the T20 sample and our large \ion{C}{4} EW quasar sample together, and the Spearman rank-order test statistic ($\rho$) and probability ($P_{\rm null}$) are reported in the bottom left.  \label{figRCIV}}
\end{figure}

Thus, only the \ion{He}{2} line enhancement is well explained by the observed
level of excess hardness of the ionizing continuum. The observed \ion{C}{4} 
excess emission is significantly stronger than predicted, while the observed 
\ion{Mg}{2} excess is significantly weaker than predicted. These
mismatches become more pronounced in the \ion{C}{4}/\ion{Mg}{2} line ratios
(last column in Table~\ref{tabRPC}). The ratio is predicted to increase 
only by a factor of 1.11, that is an excess of 0.11 in the line ratio, 
while the average observed excess is $1.02\pm 0.16$ (see also Figure~\ref{figRCIV}). 
What physical mechanism makes the \ion{C}{4}/\ion{Mg}{2} line ratio about 
twice as strong in very high \ion{C}{4} luminosity quasars?
 
\subsubsection{The effect of the gas metallicity} \label{subsubsec:z}

Since the only two free parameters that affect the line strength are the shape
of the ionizing continuum and the gas metallicity, we checked the effect of metallicity.
The third line in Table~\ref{tabRPC} shows the EWs of the four lines for the mean
ionizing continuum of $\alpha_{\rm ox}=-1.45$ and sub-Solar metallicity of $Z=0.5$ (see 
Figure~5 of \citealt{Baskin2014}). The \ion{C}{4} line is enhanced since it remains the main 
gas coolant, while the \ion{Mg}{2} and the \ion{C}{3]} lines become weaker since 
they are more minor coolants and the strength of such coolants depends on their ionic 
abundance.  The \ion{He}{2} EW is not affected by the lower $Z$ value,
as expected since this recombination line provides a rather clean measure of the 
shape of the ionizing continuum.

The large \ion{C}{4} EW quasars show a harder ionizing continuum, so we need to include 
the metallicity and the ionizing SED in the model calculations. In \citet{Baskin2014} the two 
effects are handled separately. We therefore assume the two effects are independent, and 
derive their combined effect by multiplying their individual effects (rows 4 and 5 in 
Table~\ref{tabRPC}). 
The 6th row shows the resulting predicted $\Delta \rm EW/EW$ for the four lines.
The large mismatch in the \ion{C}{4} and the \ion{Mg}{2} EW excess now disappears.
The low metallicity which enhances the \ion{C}{4} line, together with the harder 
continuum which also enhances the \ion{C}{4} line, gives a predicted 
$\Delta \rm EW/EW=2.14$ which matches well the observed value of $2.33\pm 0.14$.
The strength of the \ion{Mg}{2} line is reduced, leading to 
a predicted $\Delta \rm EW/EW=0.67$ which now matches well the observed value
of $0.61\pm 0.10$. The \ion{C}{3]} + \ion{Si}{3]} blend is now reduced to
$\Delta \rm EW/EW=-0.10$, which remains marginally consistent with the observed value
of $0.20\pm 0.10$. As noted above, the good match of the \ion{He}{2} line
remains unaffected by the reduced metallicity.
The sub-Solar metallicity also provides a good match for the observed increase in the 
\ion{C}{4} / \ion{Mg}{2} line ratio. The predicted excess in the line ratio is now 0.88, which 
matches well the observed mean excess of $1.02\pm 0.16$.  

We therefore conclude that the observed specific pattern of the UV-line enhancements
in the large \ion{C}{4} EW quasars is well explained by the combined effect of 
a hard ionizing continuum and sub-Solar metallicity. 

The sub-Solar metallicity result is consistent with 
the interpretation that these objects lie at the extreme of low 
``eigenvector 1” (EV1) objects (\citealt{Boroson1992}), as indicated
by their large \ion{C}{4} EW and excess red wing emission (see Figure~\ref{fig5}). 
The opposite extreme, of extreme high EV1 objects, such as narrow-line Seyfert 1 galaxies, 
are generally characterized by high $L/L_{\rm Edd}$ (e.g., \citealt{Boller1996,Laor2000,Boroson2002}), 
and in addition are also characterized by high metallicity, as indicated by various UV-line ratios,
such as \ion{C}{4}/\ion{N}{5} and \mbox{\ion{C}{4}/(\ion{Si}{4}+\ion{O}{4})}
\citep{Wills1999,Shemmer2004,Shin2013}. This known trend supports the 
RPC photoionization results that the large \ion{C}{4} EW quasars, which show a sign of 
low $L/L_{\rm Edd}$ (see Figure~\ref{fig6}b), have low, sub-Solar metallicity.

\subsection{The nature of the unusually strong \mbox{X-ray} emission} \label{subsec:nature}

Why are the large \ion{C}{4} EW quasars associated with excess \mbox{X-ray} emission?
As described above, strong \ion{C}{4} emission is produced when the ionizing continuum 
is harder. Is the excess \mbox{X-ray} emission then just a selection effect?
Or, is there a specific physical mechanism that enhances both the \ion{C}{4}
and the \mbox{X-ray} emission?

A hint of a possible physical mechanism is provided by the rather strong correlation 
of $\Delta\alpha_{\rm ox}$ and the \ion{C}{4} / \ion{Mg}{2} line ratio (Figure~\ref{figRCIV}).
According to the RPC photoionization-model results, the effect of $\alpha_{\rm ox}$ on the 
ratio \ion{C}{4} / \ion{Mg}{2} is small. Specifically, the mean \ion{C}{4} EW / \ion{Mg}{2} 
EW ratio of our sample is about 2.02 times that of the T20 sample, while the model 
predicts that a change 
of $\Delta \alpha_{\rm ox}=0.3$ increases the ratio by a factor of 1.11 (Table~\ref{tabRPC}),
 i.e. 0.05~dex, while the mean observed increase is 0.31~dex (Table~\ref{tabRPC}). 
As discussed above, the higher \ion{C}{4} / \ion{Mg}{2} in the strong \ion{C}{4} quasars
is well explained by a low metallicity, specifically $Z=0.5$. In fact, the distribution of
\ion{C}{4} / \ion{Mg}{2} values presented in Figure~\ref{figRCIV}, including the
extension of the T20 sample to low values, is well explained as a metallicity effect
with a minor contribution from a changing SED. The derivation above (Table \ref{tabRPC}) 
gives a line-ratio enhancement by a factor of $\log(1.88)=0.27$ for the harder ionizing
 continuum at low $Z$.
Using the results in \citet{Baskin2014} we also derive that the high-metallicity $Z=5$ 
models, with the softer ionizing continuum, decrease the line ratio by a factor of 
$\log(0.31)=-0.51$ from the  general value. The observed mean line ratio of the entire quasar sample is 
$\log(1.42)=0.15$, so the predicted range of values for $Z=0.5$--5 is 
log (\ion{C}{4} / \ion{Mg}{2})$ = -0.36$--0.42, which matches well the observed range 
(Figure~\ref{figRCIV}). This suggests that the \ion{C}{4} / \ion{Mg}{2} ratio is a fairly clean measure of the metallicity. 

Figure~\ref{figRCIV} then largely shows the effect of metallicity upon the \mbox{X-ray} 
emission. Why would higher metallicity make the coronal \mbox{X-ray} emission weaker? 
A possible answer is that higher metallicity leads to higher mass loss from the accretion disk, 
as expected for radiation-pressure wind driving. The disk mass loss passes through the corona 
above the disk, leading to mass loading of the coronal gas, which increases the plasma 
cooling rate, making the corona cooler and thus lowering also the \mbox{X-ray} emission. The 
unusually low $Z$ of the large \ion{C}{4}, extreme low EV1, quasars suppresses
the wind, and as a result lowers the mass loading, which allows the formation of an effectively 
force-free magnetized plasma. The absence of mass loading may also allow a more highly 
magnetized corona, which may increase the heating rate due to reconnection events 
(\citealt{Yuan2019,Yuan2019II}).

We note, in addition, that radio-quiet quasars with intrinsically strong \mbox{X-ray} emission are rare (e.g.,\ \citealt{Gibson2008,Pu2020}). Using the large quasar sample in \citet{Pu2020}, we estimate the fraction of \mbox{X-ray} strong quasars ($\Delta\alpha_{\rm ox}>0.2$) among radio-quiet quasars ($R < 10$) to be only 4 percent. Some studies have discussed the possibility that accretion-disk coronae contain the magnetic fields required for the launching of jets, and only when these magnetically dominated coronae are sufficiently strong can jets be launched (e.g., \citealt{Merloni2002,Zhu2020}). Under this scenario, some factor may have prevented the jet from launching, such as the topology of the magnetic field (e.g., \citealt{Livio2003,Dexter2014}), even though the coronal \mbox{X-ray} emission is strong. Theoretical studies have suggested that when external magnetic fields are too strong, an instability can develop that causes a jet to collapse, potentially forming a compact, collimated corona above the black hole (\citealt{Yuan2019,Yuan2019II}). Thus, it is possible that our quasars failed to form jets even though their coronae are strong, which separates them from the \mbox{X-ray} strong, radio-loud quasars.

\section{SUMMARY AND FUTURE WORK} \label{sec:conclusion}

In this work, we selected eight luminous quasars from the \citet{Shen2011} quasar catalog which have \ion{C}{4} EW $\geq 150$~\AA$\ $and performed Chandra snapshot observations of them. We studied the optical/UV and \mbox{X-ray} properties of these large \ion{C}{4} EW quasars and compared them to the large samples of typical quasars in T20 and T21. The main results are summarized below:
\begin{enumerate}[(i)]
\item We fit the optical spectra of the eight target quasars and find they have large \ion{C}{4} EW excess, moderate \ion{He}{2} EW and \ion{Mg}{2} EW excess, and almost no \ion{C}{3]} EW excess compared to the T20 or T21 typical quasars with similar UV luminosities. We also added two quasars from the T20 sample to our large \ion{C}{4} EW sample since they have similar \ion{C}{4} EWs and UV luminosities. See Section~\ref{subsec:largeEW}.
\item These large \ion{C}{4} EW quasars have larger $\alpha_{\rm ox}$ than the predictions from the \mbox{$\alpha_{\rm ox}$--$L_{\rm 2500}$} relation, which is as expected since both $\Delta\alpha_{\rm ox}$ and the high-ionization emission lines can indicate the amount of ionizing radiation. The $\alpha_{\rm ox}$ values of our ten quasars are also larger than expected from extrapolation of both the \mbox{$\Delta$log(\ion{C}{4} EW)--$\Delta\alpha_{\rm ox}$} and \mbox{$\Delta$log(\ion{He}{2} EW)--$\Delta\alpha_{\rm ox}$} relations derived for typical quasars (see Figure~\ref{fig4}). See Section~\ref{subsec:Xexcess}.
\item Our targets' \ion{C}{4} blueshifts are slightly smaller than for the T20 quasars with similar UV luminosities, making them locate at one extreme of the quasar locus in the \ion{C}{4} parameter space, opposite to WLQs. See Section~\ref{subsec:EW_blsft}.
\item We investigated if the cause of the strong \mbox{X-ray} emission can be revealed by our targets' \mbox{X-ray} spectra or Eddington ratios. The intrinsic absorption is small, and the median $\Gamma_{\rm eff}$ is consistent with the median value for typical quasars. The Eddington ratios of these quasars are suggestively smaller than those of the T20 quasars. See Section~\ref{subsec:Why_excess}.
\item The observed \ion{He}{2} EW increase for our quasars agrees well with the RPC prediction for a harder ionizing continuum which 
matches the observed higher $\alpha_{\rm ox}$. This RPC model under-predicts the increase in \ion{C}{4},
and over-predicts the increase in \ion{Mg}{2}. See Section~\ref{subsubsec:ionizcont}.
\item The RPC model for a harder ionizing continuum and sub-Solar ($Z=0.5$) metallicity gas 
reproduces well the observed enhancement pattern of \ion{C}{4}, \ion{He}{2}, \ion{Mg}{2}, and \ion{C}{3]} EWs.
See Section~\ref{subsubsec:z}.
\item We find that the \ion{C}{4}/\ion{Mg}{2} line ratio, in our sample and the T20 sample, is significantly correlated 
with $\Delta\alpha_{\rm ox}$ for quasars in general. In addition, the RPC model results indicate this 
line ratio is mostly set by the gas metallicity and is nearly independent of the hardness of the 
ionizing continuum. A correlation of $\Delta\alpha_{\rm ox}$ with gas metallicity is consistent with the 
known results that objects at the opposite extreme of EV1, extremely high EV1 objects, are
characterized by both weak \mbox{X-ray} emission and high metallicity gas. See Section~\ref{subsec:nature}. 
\item The underlying physical mechanism which links the gas metallicity and the coronal \mbox{X-ray} emission
is not known. A possible mechanism may be related to mass loading of the magnetized coronal gas.  
High gas metallicity may enhance radiation-pressure driven accretion-disk winds and thereby mass
loading of the corona, decreasing coronal \mbox{X-ray} emission. See Section~\ref{subsec:nature}. 
\end{enumerate}

There are several ways this work could be extended. The large \ion{C}{4} EW quasars apparently deviate from extrapolations of typical-quasar relations in some parameter spaces, but we need a larger sample to examine this behavior further. In addition, it would be valuable to obtain rest-frame optical-NUV spectra of these quasars to investigate further links between the UV-\mbox{X-ray} continuum and other optical emission lines like the Balmer lines. Simultaneous and repeated observations of \mbox{X-ray} and UV-to-optical spectra would be helpful to reveal the variation of \mbox{X-ray} emission and emission-line properties, and higher-quality \mbox{X-ray} spectra are also needed to better measure $\Gamma$ values and thereby assess Eddington ratios.

\begin{deluxetable*}{cccccccccc}
\tabletypesize{\scriptsize}
\tablewidth{\textwidth} 
\tablenum{1}
\tablecaption{Photometric and Spectroscopic Properties of Large \ion{C}{4} EW Quasars\label{tab1}}

\tablehead{
\colhead{Object Name} & \colhead{RA} & \colhead{Dec} & \colhead{Redshift} & \colhead{$M_i(z=2)$} & \colhead{log$L_{\rm 2500}$} & \colhead{$\Delta(g-i)$} & \colhead{$R$}  & \colhead{log$L_{\rm bol}$}  & \colhead{$L/L_{\rm Edd}$} \\
\colhead{(J2000)} & \colhead{(deg)} & \colhead{(deg)} & \colhead{} & \colhead{} & \colhead{} & \colhead{} & \colhead{} & \colhead{} & \colhead{} 
}
\decimalcolnumbers
\startdata
\multicolumn{10}{c}{Chandra snapshot observed quasars}   \\ \hline
$ 000510.83-092534.9 $&$ 1.295151 $&$ -9.426368 $&$ 1.866 $&$ -26.45 $&$ 30.74 $&$ 0.14 $&$ < 2.88 $&$ 46.67\pm0.02 $&$ 0.085\pm0.073 $\\
$ 090203.91+061849.6 $&$ 135.516296 $&$ 6.313801 $&$ 1.726 $&$ -26.33 $&$ 30.69 $&$ -0.03 $&$ < 2.71 $&$ 46.69\pm0.02 $&$ 0.121\pm0.034 $\\
$ 103312.84+110555.3 $&$ 158.303513 $&$ 11.098698 $&$ 2.149 $&$ -27.08 $&$ 30.99 $&$ -0.20 $&$ < 2.15 $&$ 46.81\pm0.06 $&$ 0.101\pm0.035 $\\
$ 110632.29+292314.4 $&$ 166.634537 $&$ 29.387356 $&$ 1.948 $&$ -25.77 $&$ 30.47 $&$ -0.12 $&$ < 5.23 $&$ 46.55\pm0.04 $&$ 0.074\pm0.040 $\\
$ 111208.75+085030.4 $&$ 168.036484 $&$ 8.841791 $&$ 1.875 $&$ -26.05 $&$ 30.58 $&$ -0.12 $&$ < 4.04 $&$ 46.73\pm0.02 $&$ 0.194\pm0.110 $\\
$ 115342.81+483844.4 $&$ 178.428391 $&$ 48.645687 $&$ 2.034 $&$ -27.46 $&$ 31.14 $&$ 0.02 $&$ < 1.25 $&$ 47.06\pm0.02 $&$ 0.161\pm0.078 $\\
$ 162845.49+271742.7 $&$ 247.189545 $&$ 27.295200 $&$ 1.961 $&$ -26.44 $&$ 30.73 $&$ -0.10 $&$ < 2.99 $&$ 46.69\pm0.03 $&$ 0.325\pm0.061 $\\
$ 210839.33+100413.7 $&$ 317.163910 $&$ 10.070496 $&$ 2.154 $&$ -27.02 $&$ 30.97 $&$ 0.03 $&$ < 2.08 $&$ 46.92\pm0.05 $&$ 0.082\pm0.035 $\\ \hline
\multicolumn{10}{c}{T20 quasars}   \\ \hline
$ 103215.88+574926.4 $&$ 158.066179 $&$ 57.824016 $&$ 1.835 $&$ -26.32 $&$ 30.69 $&$ 0.01 $&$ < 3.90 $&$ 46.58\pm0.02 $&$ ... $\\
$ 125929.13+600846.0 $&$ 194.871400 $&$ 60.146117 $&$ 1.869 $&$ -25.89 $&$ 30.52 $&$ 0.15 $&$ < 4.83 $&$ 46.53\pm0.06 $&$ ... $
\enddata
\tablecomments{Col.(1): Object name in the J2000 coordinate format. Cols.(2)--(3): The SDSS position in decimal degrees. Col.(4): Redshift adopted from \citet{Hewett_Wild2010}. Col.(5): Absolute $i$-band magnitude (corrected to $z = 2$; \citealt{Richards2006}). Col.(6): Logarithm of the rest-frame 2500 $\rm \AA$ monochromatic luminosity in units of erg s$^{-1}$ Hz$^{-1}$. Col.(7): Relative SDSS $g-i$ color. Col.(8): Radio-loudness parameter.  Col.(9) Logarithm of the bolometric luminosities (erg s$^{-1}$) obtained through SED fitting (see Section~\ref{subsec:Why_excess}). Col.(10) Eddington ratio (see Section~\ref{subsec:Why_excess}). }
\end{deluxetable*}

\begin{deluxetable*}{ccccccccccc}
\tabletypesize{\scriptsize}
\tablewidth{\textwidth} 
\tablenum{2}
\tablecaption{Chandra Observations and Photometric Properties of Large \ion{C}{4} EW Quasars\label{tab2}}

\tablehead{
\colhead{Object Name} & \colhead{ID} & \colhead{Date} & \colhead{Exposure Time} & \colhead{Count Rate}  &  \colhead{H/S}  & \colhead{$\Gamma_{\rm eff}$}  & \colhead{$f_{\rm 2keV}$} & \colhead{log$L_{\rm X}$}  & \colhead{$\alpha_{\rm ox}$ } & \colhead{$\Delta\alpha_{\rm ox}$ }\\
\colhead{(J2000)} & \colhead{} & \colhead{} & \colhead{(ks)} & \colhead{(0.5--7 keV)} & \colhead{}  & \colhead{} & \colhead{} & \colhead{(2--10 keV)} & \colhead{} & \colhead{}
}
\decimalcolnumbers
\startdata
\multicolumn{11}{c}{Chandra snapshot observed quasars}   \\ \hline
$ 000510.83-092534.9 $&$ 24718 $&$\rm 2021\ May\ 10 $&$ 3.969 $&$ 12.0^{+3.0}_{-2.5} $&$ 0.66^{+0.24}_{-0.18} $&$ 2.15\pm0.36 $&$ 4.24^{+0.60}_{-0.54} $&$ 45.41^{+0.12}_{-0.13} $&$ -1.22^{+0.02}_{-0.02} $&$ 0.32^{+0.02}_{-0.02} $\\
$ 090203.91+061849.6 $&$ 24721 $&$\rm 2021\ Mar\ 9 $&$ 2.896 $&$ 5.4^{+2.8}_{-2.0} $&$ 1.60^{+1.23}_{-0.70} $&$ 1.22\pm0.76 $&$ 0.76^{+0.42}_{-0.32} $&$ 44.93^{+0.17}_{-0.18} $&$ -1.51^{+0.07}_{-0.09} $&$ 0.02^{+0.07}_{-0.09} $\\
$ 103312.84+110555.3 $&$ 24715 $&$\rm 2021\ Jun\ 26 $&$ 4.356 $&$ 10.3^{+2.7}_{-2.2} $&$ 0.87^{+0.32}_{-0.23} $&$ 1.62\pm0.32 $&$ 2.87^{+0.43}_{-0.58} $&$ 45.47^{+0.11}_{-0.11} $&$ -1.34^{+0.02}_{-0.04} $&$ 0.25^{+0.02}_{-0.04} $\\
$ 110632.29+292314.4 $&$ 24716 $&$\rm 2021\ Oct\ 21 $&$ 5.720 $&$ 4.1^{+1.6}_{-1.2} $&$ 1.18^{+0.63}_{-0.43} $&$ 1.63\pm0.41 $&$ 0.79^{+0.31}_{-0.23} $&$ 44.93^{+0.14}_{-0.15} $&$ -1.38^{+0.05}_{-0.06} $&$ 0.11^{+0.05}_{-0.06} $\\
$ 111208.75+085030.4 $&$ 24717 $&$\rm 2021\ May\ 31 $&$ 4.065 $&$ 12.0^{+3.0}_{-2.4} $&$ 1.22^{+0.43}_{-0.31} $&$ 1.49\pm0.31 $&$ 1.96^{+0.51}_{-0.41} $&$ 45.36^{+0.09}_{-0.10} $&$ -1.29^{+0.04}_{-0.04} $&$ 0.23^{+0.04}_{-0.04} $\\
$ 115342.81+483844.4 $&$ 24722 $&$\rm 2021\ Jul\ 4 $&$ 2.799 $&$ 23.9^{+4.9}_{-4.2} $&$ 0.86^{+0.23}_{-0.20} $&$ 1.95\pm0.24 $&$ 6.35^{+0.68}_{-1.05} $&$ 45.78^{+0.09}_{-0.09} $&$ -1.28^{+0.02}_{-0.03} $&$ 0.34^{+0.02}_{-0.03} $\\
$ 162845.49+271742.7 $&$ 24719 $&$\rm 2021\ Feb\ 4 $&$ 4.149 $&$ 8.4^{+2.6}_{-2.0} $&$ 0.75^{+0.31}_{-0.23} $&$ 2.10\pm0.32 $&$ 2.70^{+0.50}_{-0.60} $&$ 45.30^{+0.14}_{-0.14} $&$ -1.28^{+0.03}_{-0.04} $&$ 0.26^{+0.03}_{-0.04} $\\
$ 210839.33+100413.7 $&$ 24720 $&$\rm 2021\ Apr\ 10 $&$ 3.578 $&$ 11.7^{+3.2}_{-2.6} $&$ 0.83^{+0.31}_{-0.23} $&$ 2.19\pm0.39 $&$ 3.73^{+0.58}_{-0.77} $&$ 45.54^{+0.12}_{-0.12} $&$ -1.29^{+0.02}_{-0.04} $&$ 0.30^{+0.02}_{-0.04} $\\ \hline
\multicolumn{11}{c}{T20 quasars}   \\ \hline
$ 103215.88+574926.4 $&$ 3344 $&$\rm 2002\ May\ 1 $&$ 34.206 $&$ 18.3^{+1.1}_{-1.0} $&$ 0.34^{+0.04}_{-0.04} $&$ 1.82\pm0.07 $&$ 4.05^{+0.20}_{-0.19} $&$ 45.45^{+-0.02}_{--0.02} $&$ -1.22^{+0.01}_{-0.01} $&$ 0.32^{+0.01}_{-0.01} $\\
$ 125929.13+600846.0 $&$ 13382 $&$\rm 2012\ Sep\ 1 $&$ 13.596 $&$ 4.5^{+1.0}_{-0.8} $&$ 0.60^{+0.18}_{-0.26} $&$ 1.81\pm0.22 $&$ 0.98^{+0.19}_{-0.16} $&$ 44.92^{+-0.09}_{--0.08} $&$ -1.38^{+0.03}_{-0.03} $&$ 0.12^{+0.03}_{-0.03} $
\enddata
\tablecomments{Col.(1): Object name. Col.(2): Chandra observation ID. Col.(3): Chandra observation start date. Col.(4): Effective exposure time with background flares cleaned and vignetting corrected. Col.(5): Net count rate (10$^{-3}$ s$^{-1}$). Col.(6): Ratio of the hard-band (2--7 keV) and soft-band (0.5--2 keV) counts. Col.(7): Effective power-law photon index in the 0.5--7 keV band estimated using XSPEC. Col.(8): Flux density at rest-frame 2 keV in units of 10$^{-31}$ erg cm$^{-2}$ s$^{-1}$ Hz$^{-1}$. Col.(9): Logarithm of the rest-frame 2--10 keV luminosity in units of erg s$^{-1}$, derived from $\Gamma_{\rm eff}$ and the observed-frame 2--7 keV flux. Col.(10): Observed $\alpha_{\rm ox}$. Col.(11): The luminosity-adjusted $\alpha_{\rm ox}$, $\Delta\alpha_{\rm ox}$.}
\end{deluxetable*}

\begin{deluxetable*}{cccccc}
\tabletypesize{\scriptsize}
\tablewidth{\textwidth} 
\tablenum{3}
\tablecaption{Optical Spectral Fitting Results for Large \ion{C}{4} EW Quasars\label{tab3}}

\tablehead{
\colhead{Object Name} & \colhead{MJD} & \colhead{\ion{C}{4} EW} & \colhead{\ion{C}{4} peak} & \colhead{\ion{C}{4} blueshift} & \colhead{\ion{He}{2} EW} \\
\colhead{(J2000)} & \colhead{} & \colhead{($\rm \AA$)} & \colhead{($\rm \AA$)} & \colhead{$\rm (km\ s^{-1})$} & \colhead{($\rm \AA$)} 
}
\decimalcolnumbers
\startdata
\multicolumn{6}{c}{Chandra snapshot observed quasars}   \\ \hline
\multirow{2}{*}{$000510.83-092534.9$}    &$ 52143 $&$ 158.20\pm 7.05  $&$ 1547.48 \pm 0.34 $&$ 305.52 \pm 65.07   $&$ 7.58 \pm 1.34 $\\
                                         			       &$ 56625 $&$ 93.69  \pm  4.19  $&$ 1547.13 \pm  0.21 $&$ 374.40 \pm  41.39   $&$ 6.14  \pm  0.42$ \\ \\
\multirow{2}{*}{$090203.91+061849.6$}    &$ 52649 $&$ 157.50 \pm 8.11  $&$ 1546.43\pm 0.33 $&$ 509.60  \pm  63.17   $&$ 9.09 \pm  1.41 $\\
                                     			       &$ 52674 $&$ 120.25 \pm  4.31  $&$ 1546.78 \pm  0.45 $&$ 440.57 \pm  87.46   $&$ 5.40 \pm  1.47$ \\  \\
\multirow{2}{*}{$103312.84+110555.3$}    &$ 53090 $&$ 232.84 \pm 9.68  $&$ 1544.55\pm 0.81 $&$ 872.82\pm  157.20   $&$ 13.16 \pm  1.68 $\\
                                   			       &$ 55955 $&$ 130.98 \pm  3.87  $&$ 1543.48 \pm  0.45 $&$ 1079.18\pm  87.01   $&$ 10.12 \pm 0.79 $\\ \\
\multicolumn{1}{l}{$110632.29+292314.4$} &$ 54534 $&$ 187.72 \pm 4.92  $&$ 1547.70 \pm  0.25 $&$ 263.93  \pm  48.27   $&$ 11.72 \pm  1.21$ \\  \\
\multirow{2}{*}{$111208.75+085030.4$}    &$ 54169 $&$ 179.14\pm  5.15  $&$ 1548.07 \pm  0.36 $&$ 192.39\pm  69.36   $&$ 11.32\pm  0.87$ \\
                                    			       &$ 55956 $&$ 256.91 \pm  7.49  $&$ 1549.14 \pm  0.79 $&$ -14.62  \pm  152.55  $&$ 18.70\pm 1.18$ \\  \\
\multirow{2}{*}{$115342.81+483844.4$}    &$ 52412 $&$ 150.33\pm  5.19  $&$ 1547.59 \pm 0.24 $&$ 284.87 \pm 45.70    $&$ 10.09 \pm  1.12$ \\
                                      			      &$ 56385 $&$ 93.45 \pm 0.57  $&$ 1547.59 \pm  0.16 $&$ 284.87 \pm  31.570   $&$ 5.89 \pm  0.32$ \\ \\
\multirow{2}{*}{$162845.49+271742.7$}    &$ 54553 $&$ 164.87\pm  11.48 $&$ 1546.41 \pm  0.36 $&$ 513.72 \pm  69.27   $&$ 10.61 \pm  1.87 $\\
                                      			     &$ 55751 $&$ 172.81 \pm 10.67 $&$ 1546.76\pm  0.31 $&$ 444.87 \pm  60.10    $&$ 12.36\pm  1.05$ \\  \\
\multicolumn{1}{l}{$210839.33+100413.7$} &$ 52520 $&$ 151.50 \pm 4.60   $&$ 1548.09\pm  0.40  $&$ 187.05 \pm  76.85   $&$ 9.15 \pm  1.24 $\\ \hline  
\multicolumn{6}{c}{T20 quasars}   \\ \hline
\multirow{3}{*}{$103215.88+574926.4$}    &$ 56657 $&$ 162.25\pm 4.26  $&$ 1546.85 \pm  5.39 $&$ 428.60  \pm  1042.81 $&$ 11.23 \pm  0.81$ \\
                                     				  &$ 57448 $&$ 207.07 \pm  12.08 $&$ 1548.63 \pm  3.67 $&$ 83.77  \pm  711.20   $&$ 15.29 \pm 1.47 $\\
                                   				    &$ 57452 $&$ 172.76\pm 6.90   $&$ 1551.84 \pm  2.70  $&$ -537.98 \pm  522.60   $&$ 12.07\pm 1.30$ \\ \\ 
\multicolumn{1}{l}{$125929.13+600846.0$} &$ 56447 $&$ 191.14\pm  10.64 $&$ 1546.71\pm  0.40  $&$ 455.58 \pm  78.28   $&$ 8.51 \pm  1.37$
\enddata
\tablecomments{Col.(1): Object name. Col.(2) Modified Julian date of the SDSS observations. Col.(3) Rest-frame equivalent width of the \ion{C}{4} broad emission line measured from the Gaussian models. Col.(4) Measured rest-frame peak of the \ion{C}{4} broad emission line. Col.(5) \ion{C}{4} emission line blueshift. Col.(6) Rest-frame equivalent width of the \ion{He}{2} emission line.}
\end{deluxetable*}

\begin{deluxetable*}{ccccccc}
\tabletypesize{\scriptsize}
\tablewidth{\textwidth} 
\tablenum{4}
\tablecaption{Comparison of RPC Model Predictions and Observations\label{tabRPC}}
\tablehead{\colhead{} & \colhead{\ion{C}{4}} & \colhead{\ion{He}{2}} & \colhead{\ion{Mg}{2}} & \colhead{\ion{C}{3]} (\ion{C}{3]}+\ion{Si}{3]})} & \colhead{\ion{C}{4}/\ion{Mg}{2}}}
\startdata
EW$_{Z=1,\alpha=-1.6}	$&$	53	$&$	7.4	$&$	22.5	$&$	14\ (21.4)	$&$	2.36	$ \\
EW$_{Z=1,\alpha=-1.2}	$&$	138	$&$	17.1	$&$	52.8	$&$	18\ (31.8)	$&$	2.61	$ \\
EW$_{Z=0.5,\alpha=-1.6}	$&$	64	$&$	7.4	$&$	16	$&$	8.3\ (12.9)	$&$	4.00	$  \\ \hline
EW$_{Z=1,\alpha=-1.2}/$EW$_{Z=1,\alpha=-1.6}	$&$	2.60	$&$	2.31	$&$	2.35	$&$	1.29\ (1.49)	$&$	1.11	$  \\
EW$_{Z=0.5,\alpha=-1.6}/$EW$_{Z=1,\alpha=-1.6}	$&$	1.21	$&$	1.00	$&$	0.71	$&$	0.59\ (0.60)	$&$	1.70$ \\ 
(EW$_{Z=0.5,\alpha=-1.2}-$EW$_{Z=1,\alpha=-1.6}$)/EW$_{Z=1,\alpha=-1.6}	$$^a$&$	2.14	$&$	1.31	$&$	0.67	$&$	-0.24\ (-0.10)	$&$	0.88	$ \\
Observed  $\Delta$EW/EW	&$	2.33\pm0.14	$&$	1.23\pm0.22	$&$	0.61\pm0.10	$&$	0.20\pm0.10	$&$	1.02\pm0.16	$ 
\enddata
\tablecomments{$^a$EW$_{Z=0.5,\alpha=-1.2}$/EW$_{Z=1,\alpha=-1.6}$ is estimated by directly multiplying the factors in the 4th and 5th rows.}
\end{deluxetable*}

\begin{acknowledgments}
We thank the referee for a constructive review. WNB and FZ acknowledge support from Chandra \mbox{X-ray} Center grant GO0-21080X and Penn State ACIS Instrument Team Contract SV4-74018 (issued by the Chandra \mbox{X-ray} Center, which is operated by the Smithsonian Astrophysical Observatory for and on behalf of NASA under contract NAS8-03060). AL acknowledges support by the Israel Science Foundation (grant no. 1008/18). QN acknowledges support from a UKRI Future Leaders Fellowship (grant code: MR/T020989/1). YQX acknowledges support from NSFC grants (12025303 and 11890693), the K.C. Wong Education Foundation, and the science research grants from the China Manned Space Project with NO. CMS-CSST-2021-A06.

The Chandra ACIS Team Guaranteed Time Observations (GTO) utilized were selected by the ACIS Instrument Principal Investigator, Gordon P. Garmire, currently of the Huntingdon Institute for \mbox{X-ray} Astronomy, LLC, which is under contract to the Smithsonian Astrophysical Observatory via Contract SV2-82024.

\end{acknowledgments}

\bibliography{ms}

\end{document}